\documentclass[sigconf]{acmart}

 \copyrightyear{2026}
 \acmYear{2026}
 \setcopyright{cc}
 \setcctype{by}
 \acmConference[UIST '26]{The 39th Annual ACM Symposium on User Interface Software and Technology}{November 02--05, 2026}{Detroit, MI, USA}
 \acmBooktitle{The 39th Annual ACM Symposium on User Interface Software and Technology (UIST '26), November 02--05, 2026, Detroit, MI, USA}
 \acmDOI{10.1145/3830398.3830561}
 \acmISBN{979-8-4007-2856-3/2026/11}

\usepackage{enumitem}

\usepackage{xcolor}
\usepackage[normalem]{ulem}
\usepackage{booktabs}
\usepackage{pifont}

\newif\ifshowrevisions
\showrevisionsfalse

\ifshowrevisions
  \newcommand{\rev}[1]{\textcolor{blue}{#1}}
  \newcommand{\del}[1]{\textcolor{red}{\sout{#1}}}
\else
  \newcommand{\rev}[1]{#1}
  \newcommand{\del}[1]{}
  
\fi

\begin{document}

\newcommand{\projecttitle}{Textro}
\title{\projecttitle: A Prototyping Toolkit for Solderless and Chipless Smart Textile Interfaces
 }


\author{Yanfeng Zhao}
\orcid{0009-0008-0668-0429}
\affiliation{%
  \department{Department of Computer Science}
  \institution{Florida State University}
  \city{Tallahassee}
  \state{Florida}
  \country{USA}
}
\email{yz24f@fsu.edu}

\author{Te-Yen Wu}
\orcid{0000-0003-3977-9093}
\affiliation{%
  \department{Computer Science}
  \institution{Florida State University}
  \city{Tallahassee}
  \state{Florida}
  \country{USA}}
\email{teyen.wu@fsu.edu}

\renewcommand{\shortauthors}{Yanfeng Zhao and Te-Yen Wu}

\begin{abstract}
In this paper, we present \projecttitle{}, a prototyping toolkit for designing, fabricating, and testing solderless and chipless smart textile interfaces. Unlike prior approaches that rely on rigid components or soldered connections, \projecttitle{} enables users to build functional textile interfaces using only readily available materials and tools. The toolkit integrates three parts: (1) a web-based design environment for importing sewing patterns, defining sensing elements, and automatically generating optimized component and circuit designs based on empirical experiments; (2) a fabrication pipeline that generates fabrication files for embroidery and cutting machines, with embroidery optimized for one-stroke continuous stitching paths and components assembled through glue-based attachment methods via capacitive coupling; and (3) a reader device and software for wirelessly retrieving sensor data and visualizing real-time sensor signals. We demonstrate \projecttitle{} through four application examples and conduct a user study with fashion experts, makers, and novices, highlighting its usability and potential for smart textile prototyping.
\end{abstract}



\begin{teaserfigure}
  \includegraphics[width=\textwidth]{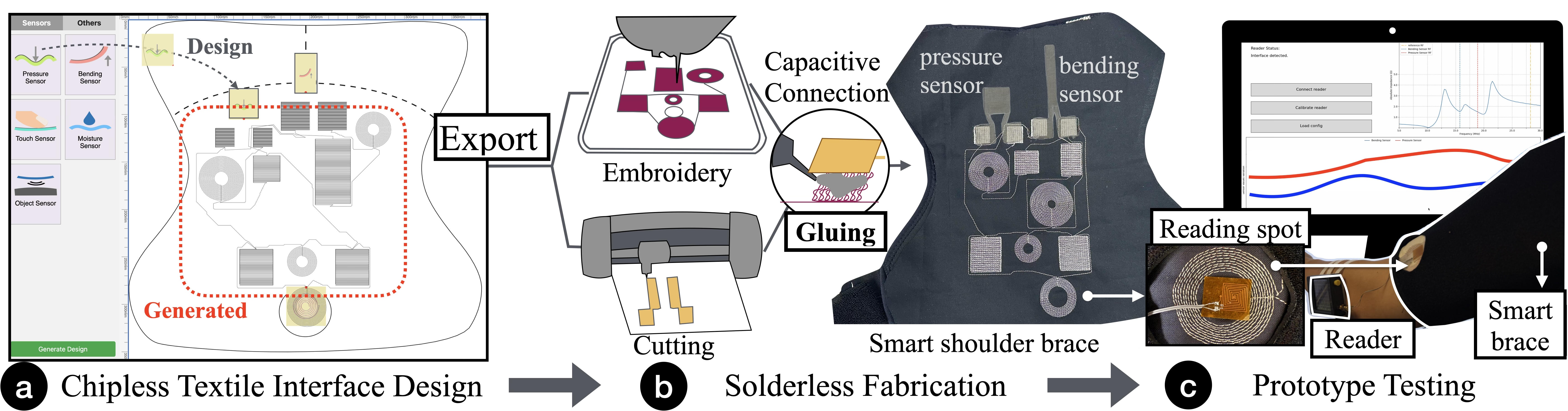}
  \caption{We introduce Textro, a solderless prototyping toolkit for chipless smart textile interfaces. (a) This toolkit features a web-based design environment, allowing users to import a garment layout (e.g., shoulder brace sewing patterns) and place sensing elements on it. (b) After finalizing the design, the toolkit generates fabrication files that guide embroidery and cutting machines to make the textile circuits and sensors, which are then assembled using fabric glue without soldering. (c) Finally, the reader and testing software enable users to validate the functionality of the prototype by wirelessly retrieving sensor data and visualizing their real time signals.}
  \Description{}
  \label{fig:teaser}
\end{teaserfigure}



\maketitle

\section{INTRODUCTION}
The growing need for seamless sensing of user input \cite{Parzer2017, Wu2020Fabriccio,2023_SmartGlove_Fan}, activity \cite{yu2024seamposerepurposingseamscapacitive, 2024_MoCap_Zhou, Wu2021, WU_IntelliLining, geissler2024embedding}, and physiological signals \cite{2022_graphelectrode_Cui, Shao2024Joey,2024_SmartTattoo_Pirrera,Lai2021Smart} across the human body has driven the rapid development of smart textile interfaces.
However, real-world adoption remains limited because embedding rigid, bulky electronics (e.g., ICs, PCBs, connectors) into soft materials increases fabrication complexity and cost \cite{Molla2017Surface, Scaling,ProForm_R1,DissolvPCB_R1}. In particular, connecting soft textiles to rigid parts often requires specialized assembly and soldering, which reduces accessibility and limits customization. Moreover, rigid components can also compromise comfort and flexibility, making it impractical to scale across the many garments people own and wear.
 
Fortunately, the emergence of battery-less and IC-less approaches offers a promising solution \cite{WU_BIT}. By integrating a coil with multi-resonant resistor–inductor–capacitor (RLC) sensing circuits into textiles, this approach enables sensing operation without embedding any batteries or circuit boards into the fabric. An external reader, equipped with its own coil, can then wirelessly couple with the textile coil and extract sensor data through spectrum scanning, further eliminating the need for rigid connectors. With this technology, and by making resistive, capacitive, and inductive components directly from textiles, it is feasible to realize fully passive and chipless smart textile interfaces, which could lower the fabrication barrier and broaden \del{accessibility and customization.} \rev{the participation in smart-textile creation among people without specialized electronics expertise, such as apparel designers and makers.}

Despite that, the democratization of this chipless approach faces several significant challenges. First, there is a knowledge gap to fabricate textile-based components with consistent and predictable electrical values using readily available tools and materials, yet such consistency is essential for system operation. Second, designing high-frequency circuits requires specialized expertise in RF engineering and circuit behavior, posing a steep barrier for non-expert users. Third, even if components and circuits can be reliably designed and fabricated, connecting them still involves soldering joints, which requires specialized tools and skills. This step not only slows down prototyping but also makes the technology inaccessible to many potential creators and makers.

To address these challenges, we present \projecttitle{}, a prototyping toolkit for designing, fabricating, and testing chipless smart textile interfaces. The toolkit features a web-based design environment that allows users to import sewing patterns and place sensing elements and a reading spot (i.e. coil) within the imported garment layout to define the intended sensing functionality (Figure 1a). Once it is complete, \projecttitle{} automatically generates optimized component and circuit designs along with machine-compatible files for fabrication. During the fabrication, these files guide embroidery and cutting machines to create the textile-based circuits and sensors, which are then assembled with fabric glue, without the need for soldering or using rigid connectors (Figure 1b). In the final testing stage, the toolkit includes a reader device and testing software that captures and shows real-time sensor signals, enabling users to quickly validate the functionality of prototypes (Figure 1c).

The development of \projecttitle{} is built upon three key advancements. \textbf{First, we study and optimize the design of textile capacitors and inductors} to ensure they can be reliably fabricated and behave with consistent and predictable electrical values. \textbf{Second, we create a solderless fabrication pipeline} that allows parallel RLC circuits in chipless smart textile interfaces to be embroidered as multiple one-stroke continuous paths, eliminating the need for manual post-fabrication connections. To further support the integration with fabric-based sensors, we also introduce a glue-based connection method that forms electrical connectivity through high-frequency capacitive coupling rather than soldering. \textbf{Third, we provide a web-based design environment that automates circuit and component generation}, enabling users to focus on sensor placement and application goals rather than low-level circuit design. 
To demonstrate the capabilities of \projecttitle{}, we develop four applications, including a shoulder brace, a knee support, a post-surgery wrap, and an interactive sling bag. Finally, we conduct a user study with 15 participants, ranging from fashion experts, makers, and novices, to understand the usability of the toolkit and gather their feedback on the workflow. 

\section{RELATED WORK}

\subsection{Chipless Interfaces}
Most chipless interfaces are fundamentally based on the principle of resonators. Resonators are materials \cite{Prakash_Gupta_2022_metamaterial, WEI202569}, patterns \cite{selfresonantcoil}, or circuits \cite{RIISTAMA2010313passivebiopotential, Galli2023Passive, niu_wireless_2019, Sun_2021_Passive} that exhibit strong frequency-selective responses, typically resonance, at specific frequencies. These resonators can be wirelessly excited and interrogated using transmitted radio signals, enabling fully passive sensing and communication without the need for embedded chips or batteries. 

In this space, circuit-based resonators are particularly promising due to their accessible materials and potential for scalable construction. These resonators consist of discrete or distributed passive components, such as resistors (R), capacitors (C) and inductors (L), arranged to generate well-defined impedance responses at specific frequencies. A typical circuit-based resonator is an LC circuit, where the inductor acts as a receiver coil for wireless access, and the capacitor functions as a sensor to detect strain \cite{Galli2023Passive, niu_wireless_2019}, pressure \cite{Sun_2021_Passive}, or fluid conductivity \cite{Charkhabi2021Monitoring}. Recent advancements, such as BIT \cite{WU_BIT}, have extended the capabilities of circuit-based resonators by enabling multiple sensors of different types, including inductive, capacitive, and resistive types, to operate concurrently with a shared receiver coil. Using a reference circuit and a mathematic model, BIT scans and interprets the impedance spectrum, supporting simultaneous readout of multiple sensing elements, enabling multimodal inputs \del{(see Apendix A for circuit schematic and operating principle)}.

Despite recent advancements, building chipless interfaces based on resonator principles remains significantly challenging. This is due to the knowledge gap of designing and fabricating passive components, such as textile-based capacitors and inductors, with predictable and consistent electrical properties required for reliable resonance. In addition, the design of high-frequency resonant circuits demands specialized knowledge in electrical engineering, which is often outside the expertise of typical users in the textile or design domains. These challenges create a steep entry barrier, limiting the accessibility and scalability of chipless interfaces. 

Our work addresses these problems with a prototyping toolkit that automates component and circuit design and uses embroidery-based fabrication with capacitive coupling to eliminate soldering. This enables users to build fully textile interfaces with \del{accessible} \rev{common }tools and materials, without requiring deep electronics expertise.
\vspace{-3mm}
\subsection{Textile-based Electronic Components}
To enable integration of electronics into fabrics, textile-based electronic components have been studied over the past decades. Various electronic components such as resistors \cite{resistor_pam, Knitted_Resistive_Fabric, Resistivity_Fabric_Lekpittaya}, capacitors \cite{blecha2017printed,inkjet_capacitor_Li, aging_passive}, inductors \cite{blecha2017printed, Roh_2010, Gong2019, aging_passive} \cite{Wu2020Fabriccio, antenna_review}, and even non-linear elements like diodes \cite{fiber_diode_woo, LED_Fiber_Hwang} can be constructed from textile materials. Among these, passive components such as resistors, capacitors, and inductors are the most commonly explored and widely used as sensors in smart textile interfaces in HCI \cite{Aigner2021, Gong2019, Parzer2018, Aigner2020, Wu2020Capacitivo, Parzer2017}~\cite{10.1145/3746059.3747733, 10.1145/3526114.3558656}. 

To accommodate different usage scenarios (e.g, different sensing purposes), prior research has explored various designs for these components. For example, textile resistors have been realized through resistive fabric \cite{Knitted_Resistive_Fabric, Resistivity_Fabric_Lekpittaya} or different lengths of conductive threads \cite{Parzer2018}  or ink \cite{Review_Conductive_Ink}. Capacitors are commonly implemented using layered structures where two conductive textile layers sandwich a dielectric layer \cite{TextileSupercapacitors, Parzer2017}. There are also diverse patterns, such as interdigitated \cite{blecha2017printed}, meandered \cite{Aigner2021} or grid patterns \cite{Wu2020Capacitivo, GoogleJacquard2021},  woven \cite{GoogleJacquard2021}, knitted \cite{Matsouka02012018}, embroidered \cite{Aigner2021} or printed onto the fabric surface \cite{blecha2017printed}. On the other hand, textile inductors are typically formed by embroidering or coiling conductive threads into spiral or meandered patterns \cite{blecha2017printed}. Their inductance is adjustable by modifying the number of turns, loop area, spacing, and layout shape. 

While these diverse designs demonstrate the feasibility of textile-based passive components, most prior work focuses on proof-of-concept implementations or requires specialized fabrication processes. There remains a lack of research on how these components can be reliably fabricated using home-grade machines, such as consumer sewing or embroidery machines, while maintaining electrical consistency and predictability. \del{Our work addresses this gap by studying and optimizing textile capacitor and inductor designs for readily available tools and materials, enabling broader adoption of chipless smart textile interfaces.}
\rev{Our work addresses this gap by studying and optimizing textile capacitor and inductor designs for common digital fabrication tools and commercially available textile materials, supporting broader adoption of chipless smart textile interface.}

\vspace{-3mm}
\subsection{E-Textile Toolkits}
To lower the barrier to wearable computing and e-textile, HCI researchers have developed various toolkits. 
The Lilypad \cite{Buechley2008, Buechley2010} is one of the earliest examples, enabling hobbyists to integrate rigid electronics into wearables and clothing.
Later toolkits such as MakerWear \cite{Kazemitabaar2017} and Brookdale  \cite{Seyed2021Rethinking} are similar tools but used the modular approaches for wearable computing construction. Beyond facilitating the integration of rigid modules, other toolkits have focused on e-textile or on-skin interface construction, using approaches such as embroidery \cite{Hamdan2018,10.1145/3672539.3686317,EMTex}, knitting \cite{luo2021knitui,laytex}, screen-printing \cite{Khan_SoftInkjet}, iron-on \cite{Klamka2020} or temporary tattoos \cite{Pourjafarian_BodyStylus} to embed sensors, actuators, and circuits into soft or flexible substrates. While these toolkits lower the floor to wearable interface creation, they still rely on rigid electronics, such as microcontrollers and batteries to operate. As a result, the minimum required technical and fabrication effort remains high, hindering broader adoption and customization.

In contrast to existing toolkits, \projecttitle{} is designed to support chipless interfaces that operate using high-frequency signals. This enables unique advantages, such as eliminating the integration of rigid components and allowing solderless and wire-free interconnects. However, these benefits come with new challenges, such as designing and fabricating those high-frequency textile-based components and circuits. We address these challenges by abstracting component designs through empirical studies to parameterize textile components for predictable electrical behavior, and by automating circuit design and layout generation to ensure direct fabrication with embroidery and cutting machines.

\section{OPERATING PRINCIPLE OF CHIPLESS SMART TEXTILE INTERFACES}
\rev{This section explains the working principle of chipless smart textile interfaces and the circuit structure required for wireless readout.}

\begin{figure}[h!]
    \centering
    \includegraphics[width=1\linewidth]{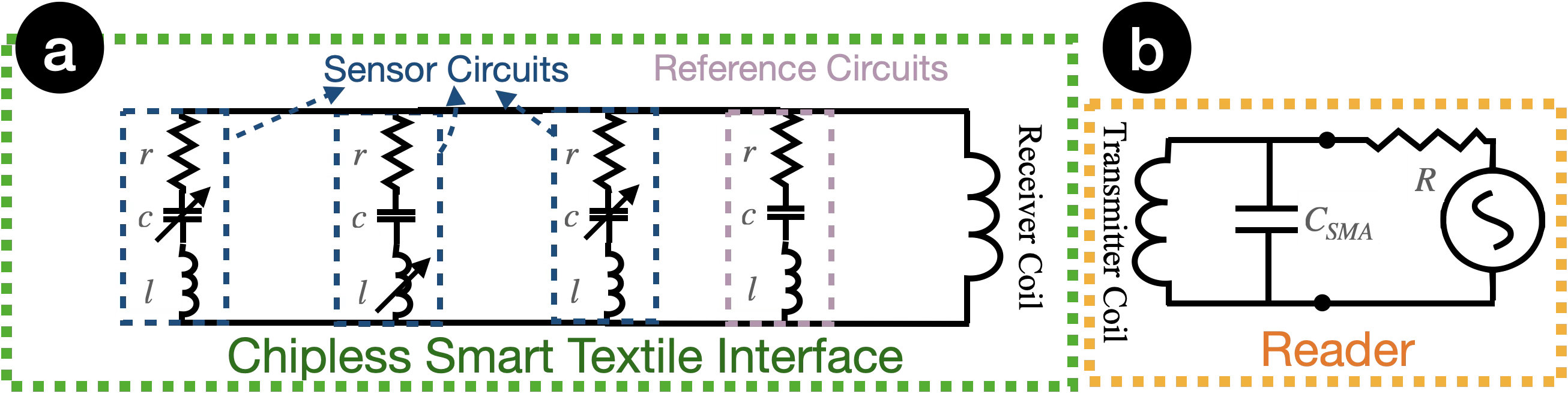}
    \caption{Circuit structure and operating principle of chipless smart textile interfaces. 
    (a) The textile interface consists of multiple sensor RLC circuits and a reference RLC circuit connected in parallel with a receiver coil. 
    (b) The external reader uses a transmitter coil to wirelessly couple with the receiver coil and scan the frequency-domain response for sensor readout.}
    \label{fig:operating_principle}
    \vspace{-2mm}
\end{figure}

\rev{In light of the interface design proposed by BIT~\cite{WU_BIT}, a chipless smart textile interface consists of a receiver coil connected in parallel with a reference circuit and multiple sensor circuits, as shown in Figure~\ref{fig:operating_principle}a. Both the reference and sensor circuits are resistor--inductor--capacitor (RLC) resonators whose inductance and capacitance values determine their frequency response. The reference circuit provides a stable baseline that helps calibrate the coil coupling and locate the resonant frequencies of the sensor circuits.
In each sensor circuit, the sensing component can replace the resistor, inductor, or capacitor in the RLC circuit, allowing user interactions or environmental changes to modulate the circuit impedance. By assigning each sensor circuit to a distinct frequency range, multiple sensors can be distinguished through frequency-domain analysis. During readout, an external reader with a transmitter coil wirelessly couples with the receiver coil and scans the impedance spectrum of the textile interface (Figure~\ref{fig:operating_principle}b). Using a mathematical model and the known reference resonance, the system decodes the measured spectrum to estimate the values of all connected sensors.}

\section{TEXTRO WALKTHROUGH}\label{sec:walkthrough}

{The goal of \projecttitle{} is to empower users to easily create smart textile interfaces using common tools and materials, without the need for embedding and soldering rigid electronics, batteries, and connectors into textiles. This section demonstrates a running example to illustrate the workflow of using \projecttitle{} to design, fabricate and test a chipless smart textile interface (see Video Figure). The example is a smart shoulder support brace where textile-based bending and pressure sensors are integrated into the shoulder area to monitor joint movement and rotation.

\del{Ben is a physical therapist. He wants to use \projecttitle{} to create a personalized smart shoulder brace, helping a patient monitor shoulder mobility and recovery progress.}
\rev{Ben is an apparel designer collaborating with a physical therapist to create a personalized smart shoulder brace for monitoring shoulder mobility and recovery progress.}
\rev{He begins by uploading a sewing pattern of the brace into the web-based design environment (Figure~\ref{fig:walkthrough_1}a). He places a bending sensor at the shoulder joint, a pressure sensor over the acromion, and a reading spot on the upper arm where an external reader can access the textile interface (Figure~\ref{fig:walkthrough_1}b--c). Once satisfied with the placement, Ben generates the interface design, previews its alignment with the sewing pattern, and downloads the fabrication and configuration files (Figure~\ref{fig:walkthrough_1}d--e).
}


\begin{figure}[h]
    \centering
    \includegraphics[width=1\linewidth]{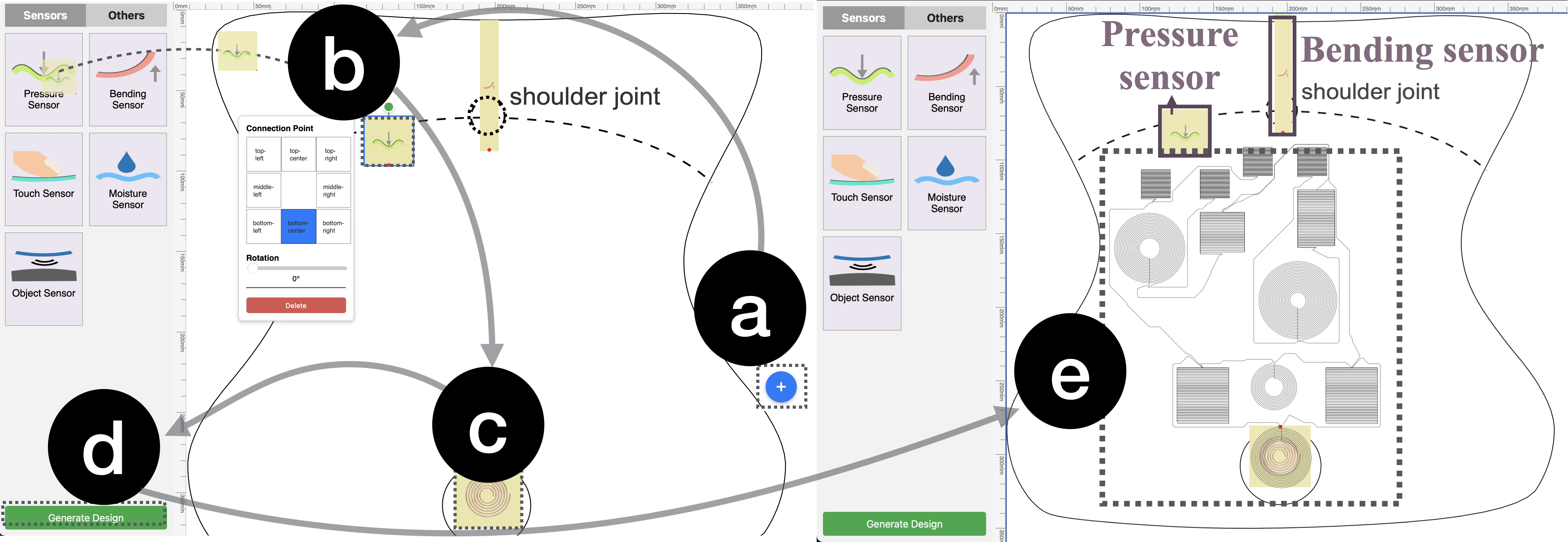}
    \caption{Walkthrough of \projecttitle{}’s design environment. The workflow begins by (a) adding a sewing pattern, followed by (b) placing and configuring sensors at desired locations. Users can then (c) add a reading spot, (d) generate the interface design, and (e) preview the resulting layout.}
    \label{fig:walkthrough_1}

\end{figure}

\rev{Ben then fabricates the embroidered circuits and textile sensors using the generated embroidery and cutting files (Figure~\ref{fig:walkthrough_2}a--b). He attaches the fabric-based sensors to the embroidered connection electrodes with fabric glue, forming solderless capacitive couplings between the sensors and the circuit (Figure~\ref{fig:walkthrough_2}c). Finally, he cuts the assembled interface according to the sewing pattern and stitches it into the shoulder brace.}

%

\begin{figure}[h]
    \centering
    \includegraphics[width=1\linewidth]{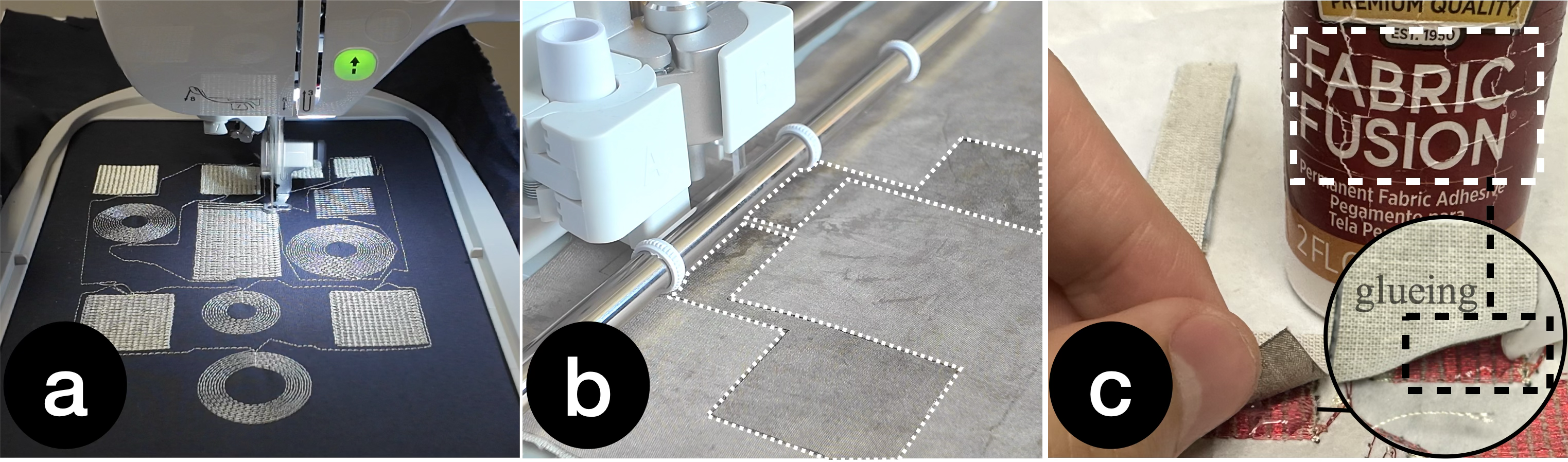}
    \caption{\projecttitle{}'s fabrication workflow. (a) Essential components and circuits directly embroidered onto fabric using a sewing machine. (b) Fabric-based sensors with their electrodes are made using a desktop cutter. (c) Sensors attached to the embroidered circuits via fabric fusion glue.}
    \label{fig:walkthrough_2}
\end{figure}

\rev{To test the prototype, Ben connects the reader to the testing software, loads the generated configuration file, and places the reader coil on the designated reading spot (Figure~\ref{fig:walkthrough_3}a--c). The software then visualizes sensor values in real time, allowing Ben to observe changes as the shoulder moves and to verify that the smart brace prototype is functional (Figure~\ref{fig:walkthrough_3}d).}

\begin{figure}[h!]
    \centering
    \includegraphics[width=1\linewidth]{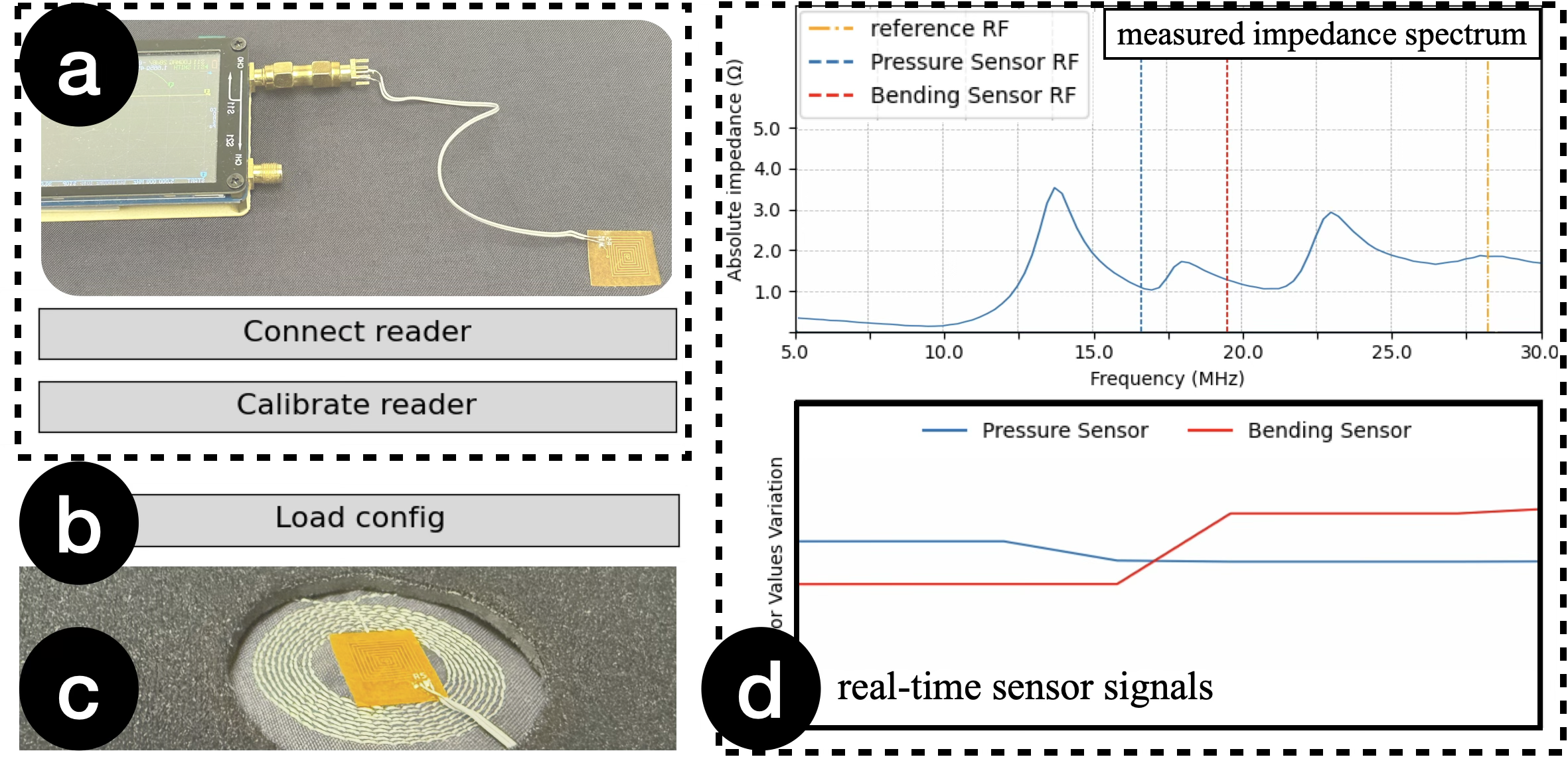}
    \caption{\projecttitle{}'s testing workflow. (a) Connect and calibrate the reader. (b) Upload the configuration file. (c) Position the reader’s coil on the designated reading area. (d) Observe sensor value changes during interactions.}
    \label{fig:walkthrough_3}
    \vspace{-3mm}
\end{figure}

\section{TEXTILE COMPONENT DESIGN}

A key challenge in Textro lies in designing textile capacitors and inductors whose component values remain consistent across fabrication and use. Without such consistency, the resulting resonant circuits may behave unreliably. Furthermore, if their geometric parameters, such as width and length, cannot be systematically mapped to predictable component values, Textro cannot support the computational design of chipless smart textile interfaces. 

To address this challenge, \rev{this section has two goals: (1) to identify textile capacitor and inductor designs that remain robust across fabrication, post-laundry, and on-body conditions; and (2) to derive empirical models that map geometric parameters to capacitance and inductance values, enabling \projecttitle{} to generate textile components with target electrical values.}

\vspace{-2mm}
\subsection{Textile Capacitor}
Among many textile capacitor designs, we select and evaluate those that are reported to balance high capacitance density with low-cost, accessible fabrication \cite{Aigner2021, InkjetCapacitors, aging_passive}: interdigital, meandered spiral, and parallel-plate. We fabricate these designs using three methods, 34~AWG embroidery with enamel-coated wire~\cite{BNTECHGO2022}, 34~AWG embroidery with bare conductive wire~\cite{Belden8057}, and cut-and-transfer using conductive fabric~\cite{ConductiveFabric}, resulting in nine capacitor types in total. Detailed fabrication parameters are provided in Appendix~\ref{Capacitor Fabrication Parameters}.

To quantify robustness, we use a Vector Network Analyzer (NanoVNA) to measure capacitance variation from three sources of error: fabrication, laundry, and on-body use. Fabrication error captures variability across five replicated samples caused by inconsistencies in stitching or cutting. On-body error measures variation across five replicates when the capacitors are wrapped on the arm with the textile substrate. Post-laundry error captures capacitance changes after 10 wash-and-dry cycles, with samples placed in a meshed wash bag. Each cycle consists of 48 minutes of delicate-mode washing with detergent, followed by 90 minutes of delicate-mode drying. Detailed procedures are provided in Appendix~\ref{Capacitor Evaluation Procedures}.


\begin{figure}[b]
    \vspace{-2mm}
    \centering
    \includegraphics[width=1\linewidth]{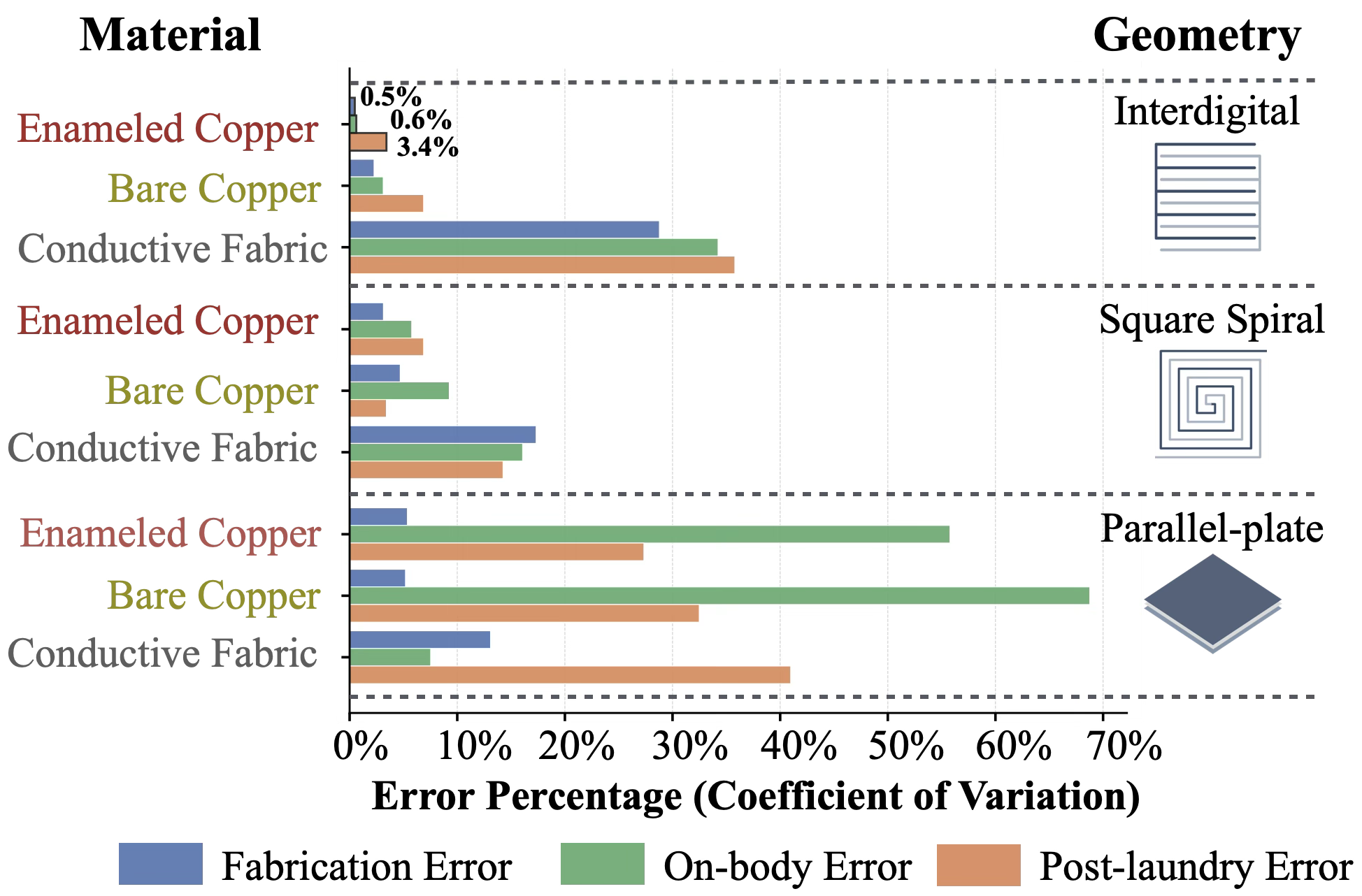}
    \caption{Experimental results on fabrication, post-laundry, and on-body error for each design. The interdigital capacitor embroidered with enameled copper wire achieved the best overall performance.}
    \label{fig:capacitor_design_result}
    \vspace{-2mm}
\end{figure}
The experimental results are shown in Figure~\ref{fig:capacitor_design_result}. 
The interdigital capacitor embroidered with enameled wire exhibits the lowest variation across fabrication, post-laundry, and on-body conditions, with coefficients of variation of 0.5\%, 0.6\%, and 3.4\%, respectively. 
This robustness likely stems from the small spacing between interdigital fingers, which helps maintain a consistent electric field distribution, and from the enamel coating, which insulates conductive paths and reduces the risk of short circuits during laundry or skin contact. In contrast, interdigital capacitors made with bare wire or conductive fabric show substantially higher variability due to greater susceptibility to short circuits and human body interference. The meandered spiral design also performs relatively well, but its cornered geometry introduces greater fabrication variability. The parallel-plate design shows the highest instability because its capacitance is highly sensitive to changes in the gap between conductive layers.

Based on these results, we select the interdigital capacitor embroidered with enameled wire as the final design. We then derive its empirical formulas to predict the capacitance value based on geometric parameters. To ground our formula in established theory, we begin with the standard PCB interdigital capacitor formula \cite{pozar_microwave_2012} and refine it using empirical measurements from 100 textile prototypes (i.e., $(10\ \text{finger lengths} + 10\ \text{finger counts}) \times 5$ replicates). Because capacitance differs between off-body and on-body conditions, we derive separate formulas. For the on-body condition, the prototypes are measured while wrapped around the arm with the 1mm thick cotton substrate. As shown in Figure~\ref{fig:capacitor_fitting}, both fitted models show high agreement with the measured data ($R^2 = 0.99$ and $R^2 = 0.95$). The detailed experimental procedures and results, including tests on different substrates, are provided in Appendix~\ref{model fitting for Interdigital Capacitor}.  \rev{In general, the models also generalized well across materials, yielding relative prediction errors below 3\% for six commonly used textile substrates: silk, polyester, cotton, nylon, linen, and muslin.}

\begin{figure}[h!]
    \vspace{-2mm}
    \centering
    \includegraphics[width=1\linewidth]{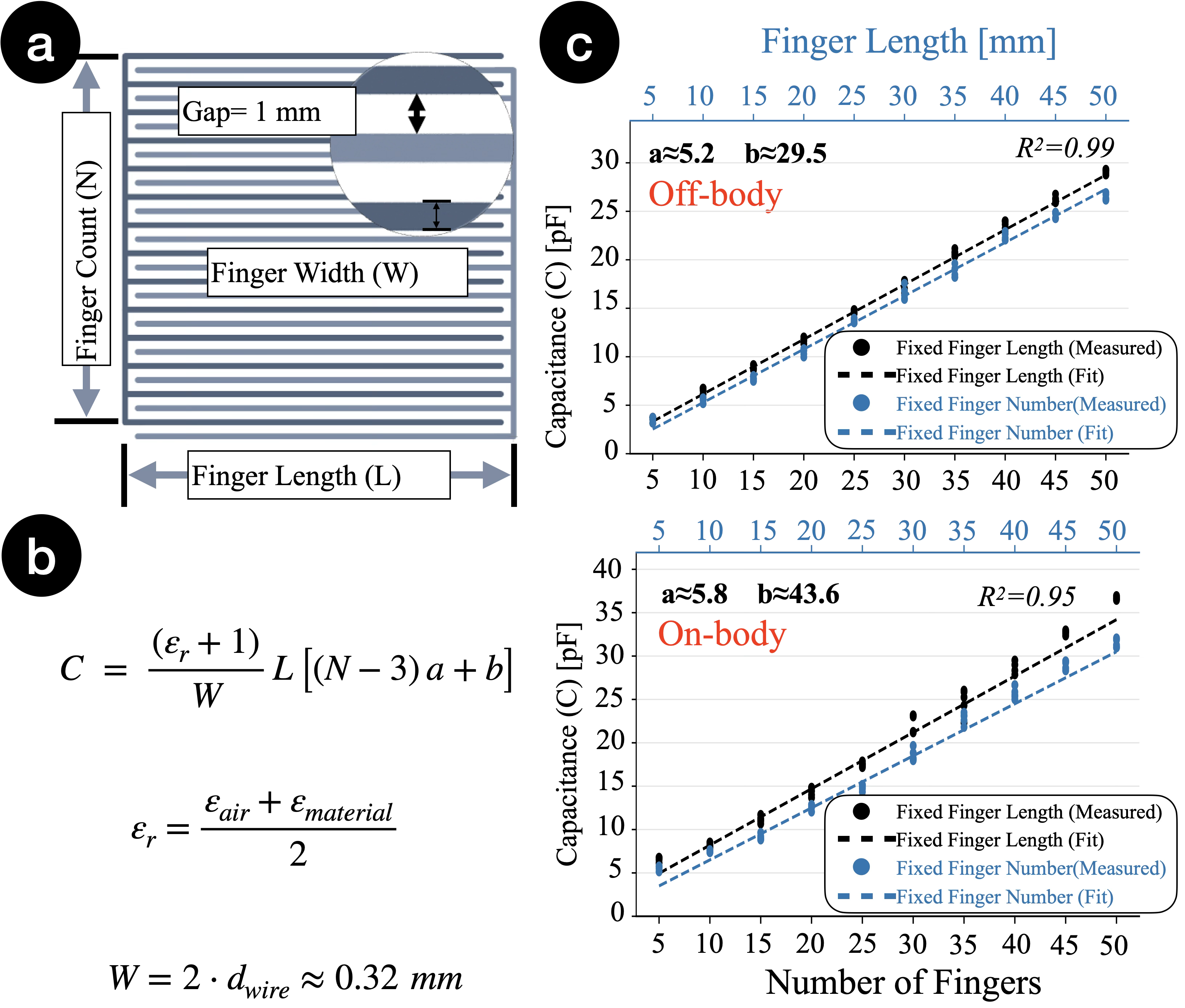}
    \caption{(a) Design parameters for interdigital capacitor. (b)Theoretical PCB interdigital capacitor fitting model. (c) Measured and fitted capacitance values under off-body~(top) and on-body~(bottom) conditions. The blue and black data points and fitted line correspond to the top~(Finger length) and bottom~(Number of Fingers) x-axes respectively.}
    \label{fig:capacitor_fitting}
    \vspace{-2mm}
\end{figure}


\subsection{Textile Inductor}
We explore three textile inductor designs with high inductance density: circular spiral, square spiral, and meander-line inductors. Similar to the textile capacitors, we fabricate these designs using accessible textile fabrication methods, including embroidery with 34~AWG enamel-coated~\cite{BNTECHGO2022}, 34~AWG bare conductive wire~\cite{Belden8057}, and cut-and-transfer using conductive fabric~\cite{ConductiveFabric}. However, spiral inductors require overlapping or intersecting paths, especially at the center where the inner trace must exit the spiral, making insulation necessary to prevent short circuits. As a result, the circular and square spiral designs can only be fabricated using embroidery with enamel-coated wire, providing the required insulation. In contrast, the meander-line design can be fabricated using all three methods. Detailed fabrication parameters are provided in Appendix~\ref{Inductor Fabrication Parameters}. We also use the same measurement procedure as for the capacitors to quantify design robustness in terms of variation from fabrication, post-laundry, and on-body use.

\begin{figure}[h]
    \vspace{-3mm}
    \centering
    \includegraphics[width=1\linewidth]{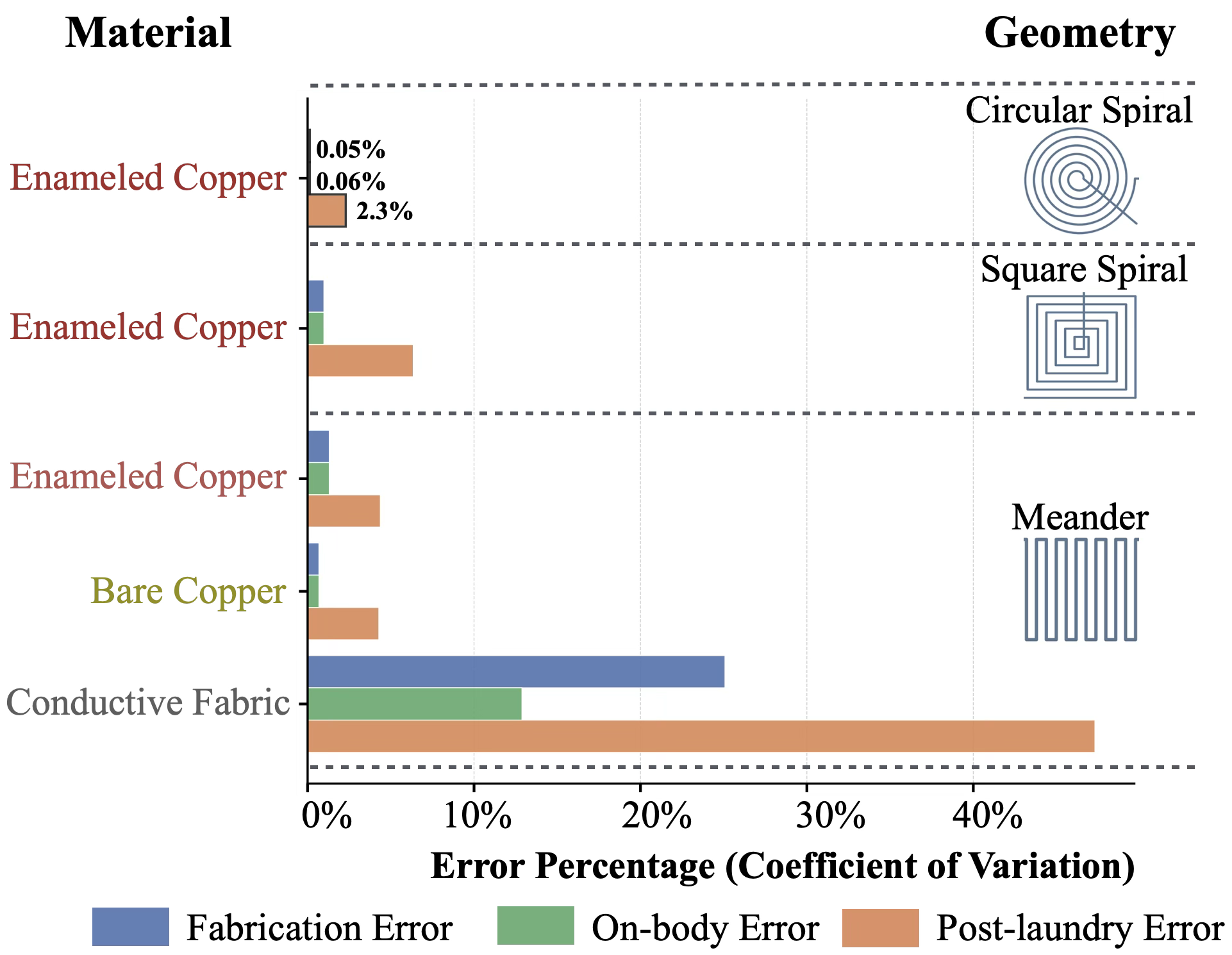}
    \caption{Experimental results showing fabrication, post-laundry, and on-body error for each design. Square and circular spirals achieved the best overall performance.}
    \label{fig:inductor_design_result}
\end{figure}

We show the experiment results in Figure \ref{fig:inductor_design_result}.  Overall, all embroidered inductor designs exhibit relatively low error rates ($<10\%$) across fabrication, post-laundry, and on-body conditions, suggesting embroidery is a reliable method for constructing textile inductors. In contrast, the cut-and-transfer approach demonstrates substantially higher variation across all error sources ($\geq5\%$). This can be attributed to the difficulty in preserving the precise spacing and shape of the meandered pattern during the manual transfer process.
Between the square spiral and circular spiral inductor designs, we select the circular spiral for our final implementation because it exhibits lower post-laundry error (circular: 2.3\%, square: 6.3\%). This difference is likely due to the circular geometry’s smoother continuous path, which better preserves trace spacing and reduces stress concentration during washing and drying, whereas the square spiral’s sharp corners are more susceptible to deformation.

Next, we derive empirical formulas for the selected textile inductor design. We begin with the standard circular spiral inductor formula \cite{Wheeler} and calibrate it using measurements from 55 textile prototypes, consisting of $(5\ \text{inner diameters} + 6\ \text{turn counts}) \times 5$ replicates. As presented in Figure \ref{fig:inductance_fitting}, the fitted model achieves strong agreement with the measured data ($R^2 = 0.99$). 
The detailed procedure is described in Appendix~\ref{model fitting for circular inductor}. Note that unlike capacitors, inductors are less influenced by the human body because their behavior is primarily determined by the magnetic field generated by the coil geometry, which is much less affected by nearby body contact.

\begin{figure}[h!]
    \centering
    \includegraphics[width=1\linewidth]{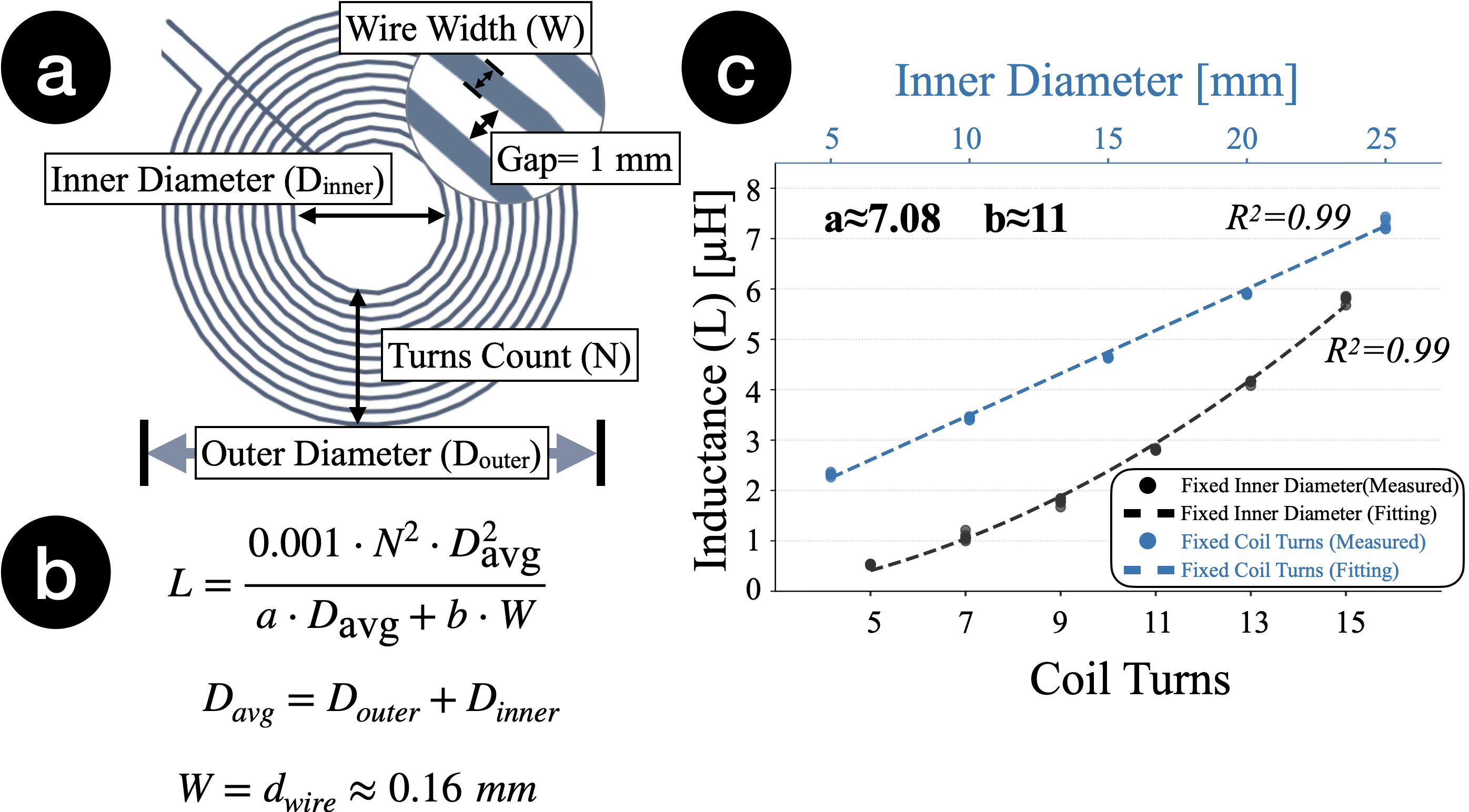}
    \caption{(a)Design parameters for circular spiral inductor. (b) Theoretical PCB circular spiral coil fitting model. (c)Empirical fitting of the circular spiral inductor model, yielding a = 7.08 and b = 11 ($R^2$ = 0.99).}
    \label{fig:inductance_fitting}
    \vspace{-4mm}
\end{figure}


\subsection{Textile Sensor}
 In this work, we focus on five representative textile sensors from prior research:\rev{1) \textit{touch sensing}, in which capacitance increases when a finger approaches or contacts the electrodes~\cite{10.1145/3654777.3676344,10.1145/3379337.3415886}; 2) \textit{moisture sensing}, in which absorbed liquid increases capacitance by changing the dielectric properties of the textile substrate~\cite{6494581}; 3) \textit{pressure sensing}, in which capacitance increases as a soft dielectric is compressed between conductive fabric electrodes~\cite{9580528,https://doi.org/10.1002/admt.201700237}; 4) \textit{bending sensing}, in which deformation changes capacitance by altering the overlap or spacing between electrode layers~\cite{geissler2024embedding}; and 5) \textit{metal-object sensing}, in which a nearby metal object decreases the inductance of an embroidered coil by perturbing its magnetic field~\cite{Gong2019}. The first four sensors operate within the 4--25~pF range in our prototypes, while the metal-object sensor operates within the 1--10~$\mu$H range. Figure~\ref{fig:sensor_design} summarizes their structures, design parameters, and measured component values.} Additional material details and potential applications are provided in Appendix~\ref{textile sensor design}.
\rev{Note that resistive sensing could also modulate the RLC response in principle, but we do not include it in this work because many commercially available soft resistive materials have relatively high resistance, which can overly damp the resonant response.} 

\begin{figure}[t]
    \centering
    \includegraphics[width=1\linewidth]{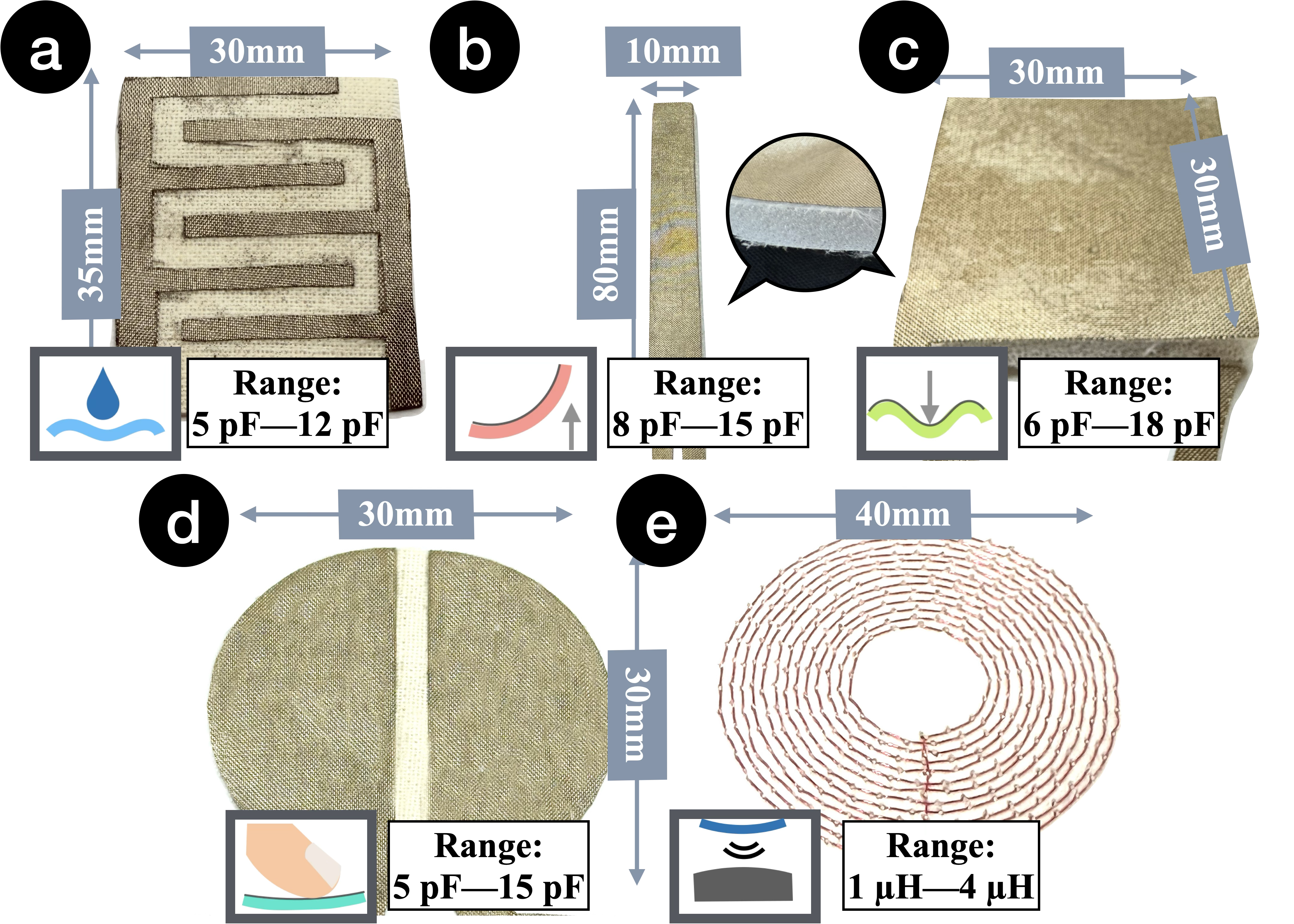}
    \caption{Five types of sensors supported by \projecttitle: (a) touch sensor, (b) moisture sensor, (c) pressure sensor, (d) bending sensor, and (e) metal object detector. Note that our capacitive sensors are implemented using woven conductive fabric due to its softer texture and comfort in sensitive areas.}
    \label{fig:sensor_design}
    \vspace{-2mm}
\end{figure}

\section{SOLDERLESS SMART TEXTILE INTERFACE DESIGN}

While our experimental results enable Textro to use these textile component designs and formulas for computational resonant circuit design, a remaining practical challenge is how to connect the textile components. Our textile capacitors and inductors are fabricated with enameled wire, which typically requires stripping the insulation and soldering wires together, steps that demand tools and skills beyond those of typical non-expert users. To address this challenge, we design each chipless smart textile interface as a set of single-path traces, allowing the entire circuit layout to be embroidered in one process without manual connection, wiring, or soldering. In addition, to incorporate textile sensors made from conductive fabric, we introduce a capacitive coupling mechanism that interfaces the sensor’s conductive surface with embroidered electrodes through stacking and gluing, eliminating the need for direct electrical contact. As a result, sensors can be attached or replaced easily without specialized tools.

\subsection{One-stroke Embroidery Layout}
The circuit of a chipless smart textile interface is a parallel circuit where multiple RLC circuits share the same receiver coil connection. In this configuration, each RLC circuit’s junction with the receiver coil forms a three-way branch: one path to the receiver coil, one to the RLC circuit itself, and one to the next RLC circuit. For an embroidery machine, this type of junction cannot be stitched in a single continuous “one-stroke” path because the machine can only proceed from one branch to another sequentially. For instance, if the machine stitches from the receiver coil into one RLC circuit, it must complete the entire RLC loop before returning to the junction. But in this way, when stitching the next RLC circuit, the entire circuit topology has changed, becoming a series connection rather than the intended parallel configuration (Figure \ref{fig:one_path_circuit}a). To connect all three paths at the junction correctly, the embroidery process would need to stop at one branch, cut the thread, and then restart at another. This introduces breaks in conductivity. When using enameled wire, restoring these connections requires manually stripping insulation and soldering the wires together, a slow, skill-intensive process that is impractical for non-expert users.

To address this challenge, we redesign the circuit of a chipless smart textile by breaking it into multiple single-stroke segments, each of which can be embroidered in continuous stitches.  The key to this redesign lies in the textile capacitor design. A capacitor inherently consists of two separate conductive electrodes that do not need to be stitched in a single continuous path. This property allows us to use capacitors as intentional “break points” in the embroidery sequence.

\begin{figure}[h!]
    \centering
    \includegraphics[width=1\linewidth]{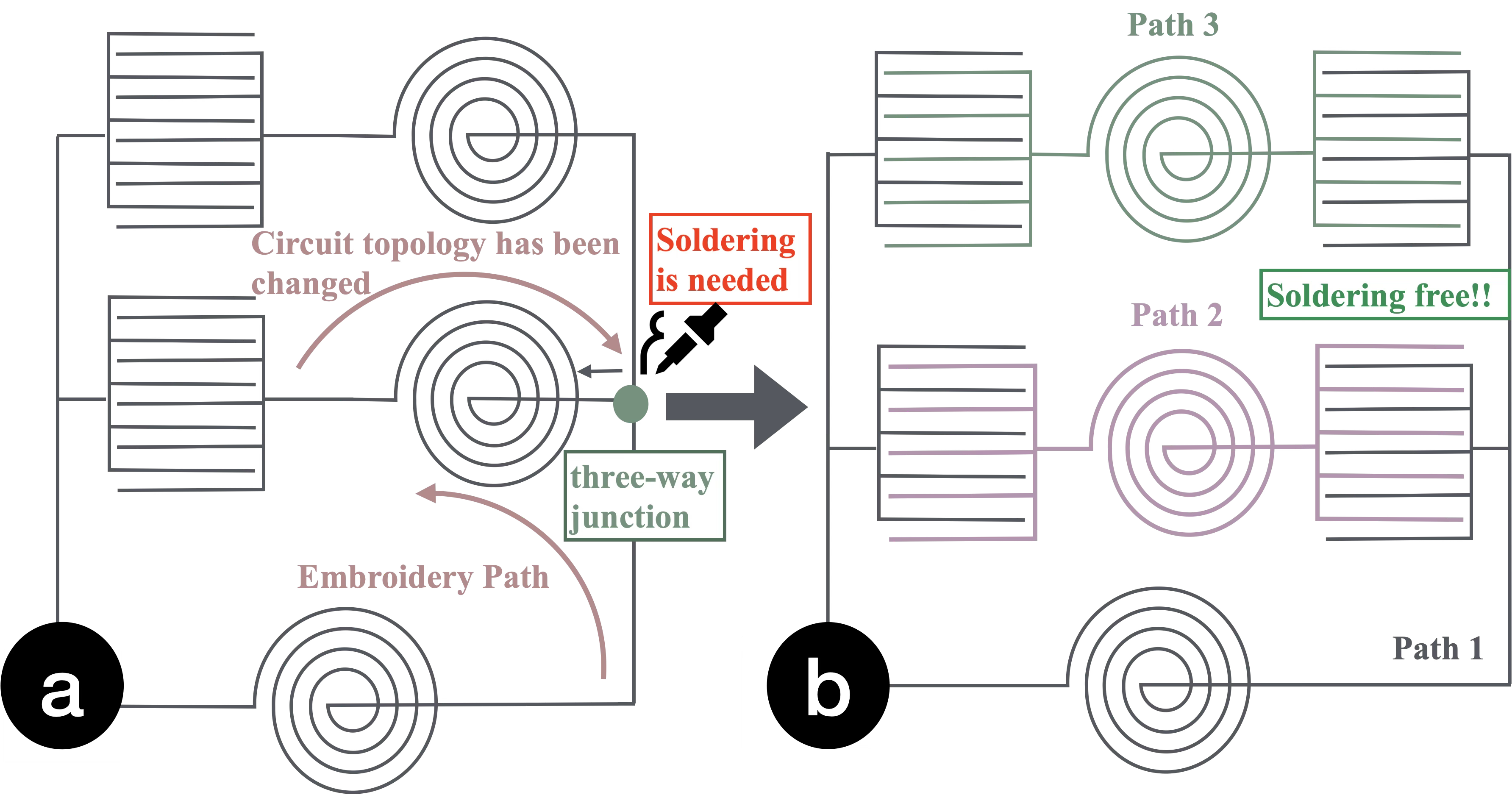}
    \caption{One-stroke embroidery challenge and solution. (a) A direct embroidery of parallel RLC circuits results in the change of circuit topology because the embroidery machine cannot form true three-way stitched junctions in a single continuous path. (b) Our redesigned layout introduces capacitors as intentional break points at each junction. This allows the receiver coil and RLC circuits to be embroidered in separate continuous paths, eliminating the need for manual thread cutting, stripping, or soldering while preserving the parallel configuration.}
    \label{fig:one_path_circuit}
\end{figure}

In our design, each junction between the receiver coil and an RLC circuit is replaced with a capacitor (Figure \ref{fig:one_path_circuit}b). During the first embroidery pass, we stitch the receiver coil and the connection to only one electrode of each capacitor at the junctions. This means the embroidery can move from junction to junction without having to complete the entire RLC loop, avoiding the problem of forming series connections. Once the receiver coil path is complete, each RLC circuit is embroidered separately in its own single continuous path, connecting to the other electrode of its capacitors.

This design allows the entire interface to be fabricated without creating any true three-way stitched junctions. Although this approach requires two capacitors for each RLC circuit, one at each end, it ensures that every segment can be stitched in a one-stroke manner. As a result, the entire chipless smart textile interface can be created directly on an embroidery machine, without the need for manual reconnection, insulation stripping, or soldering.

\begin{figure}[h!]
    \centering
    \includegraphics[width=1\linewidth]{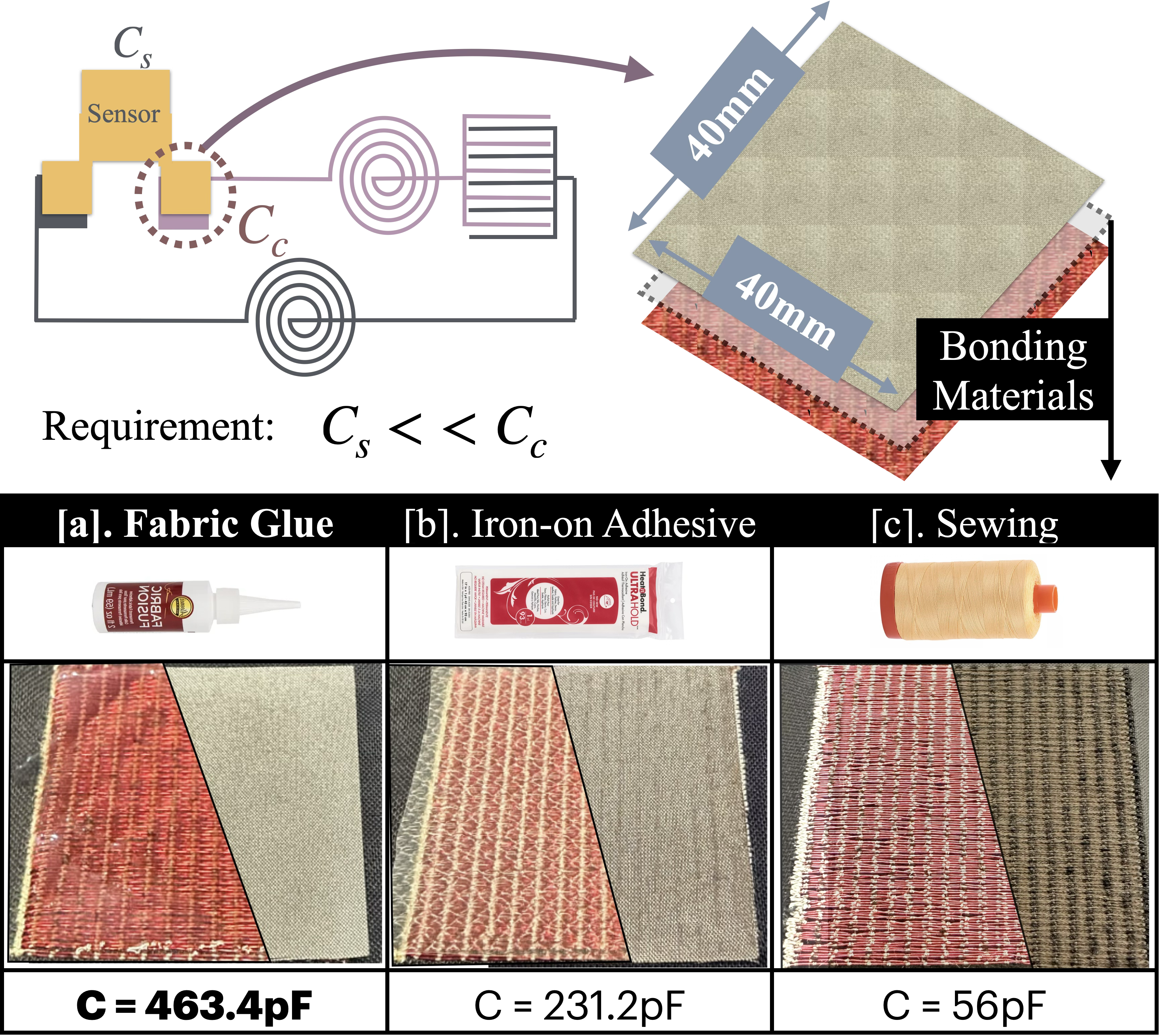}
    \caption{Capacitive coupling connection, made with three bonding methods: (a) fabric fusion glue \cite{fusion}, (b) iron-on adhesive \cite{iron-on}, and (c) sewing. The embroidered electrode is a dense comb-like pattern, fabricated using an AWG 34 enameled copper wire (Remington Industries) on a cotton substrate.}
    \label{fig:coupling_medium}
    \vspace{-2mm}
\end{figure}

\subsection{Capacitive Connection to Fabric-based Sensor}

While our redesigned interface circuit resolves the connection challenges between embroidered components, it does not address how to connect to our textile sensors. Our sensors are mostly made from conductive fabric, which introduces heterogeneity in both materials and layer structures. Creating an electrical connection between fabric-based sensors and the embroidered circuitry, without using wires or soldering, remains a challenge.

To solve this challenge, our core idea, once again, is to leverage the concept of the capacitor. By using capacitive coupling at each connection end of the textile sensor, we can avoid direct electrical connection. To integrate this mechanism into the one-stroke embroidery layout, we modify the circuit design by replacing one side of the capacitor with two embroidered coupling electrodes. One electrode is stitched as part of the main embroidery path, while the other is integrated with the sensor’s circuit, as shown in Figure \ref{fig:coupling_medium}. When the sensor's conductive fabric is stacked onto the embroidered base, the aligned electrodes form capacitive couplings that electrically interface the sensor with the rest of the circuit without soldering or wiring. The key requirement of this mechanism is the capacitive connection must have significantly larger capacitance ($C_c$) than that of the sensor itself ($C_s$). Since the two coupling capacitors and the sensor are connected in series, the smallest capacitance in the chain will dominate the overall equivalent capacitance. In this way, if the coupling capacitors are much larger than the sensor’s capacitance, their effect on the total capacitance will be minimal. According to the calculation in Appendix E, a coupling capacitance at least five times greater than the sensor’s capacitance is sufficient to maintain measurement error within 10\%.

To achieve this, we need to first maximize the coupling capacitance. As the parallel-plate design has the highest capacitance density \cite{InkjetCapacitors}, we mainly focus on that design. In our case, the parallel-plate capacitor is formed with one plate made from the sensor’s conductive fabric and the other from an embroidered electrode. The dielectric layer in between can either be the enamel coating of the embroidered electrode or an additional bonding material. The capacitance of this structure is determined by three main factors: the overlapping plate area, the separation (gap) between plates, and the permittivity of the dielectric material. Among these, the gap is the most difficult to control in fabrication and is a major source of variability. Therefore, to achieve a high capacitance without excessively increasing the plate area, we prioritize selecting a bonding material with high permittivity as the dielectric layer.

\subsubsection{Experiments on Bonding Materials}
We test three types of bonding materials: fabric fusion glue (Figure \ref{fig:coupling_medium}a), iron-on adhesive (Figure \ref{fig:coupling_medium}b), and sewing in which the enamel coating of the embroidered electrode acts as the dielectric layer (Figure \ref{fig:coupling_medium}c). For each method, we fabricate five 40 mm × 40 mm prototypes and measure their capacitance using a NanoVNA after three days of natural drying.
Our results show that the fabric glue is the most promising bonding method. Even when fully dried, the coupling capacitor made with fabric glue exhibit an average capacitance of 463.4 pF (variation= 10\%), which is significantly higher than that of the iron-on adhesive (231.2 pF, variation= 11\%) and the sewn configuration (56 pF, variation= 12\%). We attribute this to the higher permittivity of the glue compared to the other dielectrics, which allows for greater capacitance. This makes fabric glue an optimal choice for achieving both strong mechanical bonding and high capacitance in our capacitive coupling design.

\subsubsection{Experiments on Size of Capacitive Coupling Connection}
Next, we examine how the size of the capacitive coupling connection affects measured capacitance, enabling practical guidelines for matching fabric-based sensors with appropriately sized coupling connections.
In our experiment, we tested square coupling connections with side lengths ranging from 10 mm to 40 mm, in increments of 10 mm, with five replicates for each size. The results show that 10 mm × 10 mm connections yield very low capacitance (on average 18.8 pF), making them unsuitable for reliable connection. In contrast, the 20 mm × 20 mm and 30 mm × 30 mm connections achieved average capacitances of 133.8 pF (variation = 10\%) and 310.2 pF (variation = 15\%), respectively. These values are sufficiently high for most of our fabric-based sensors. 

\section{TOOLKIT DESIGN AND IMPLEMENTATION}
Building on the textile component and interface designs, we develop a toolkit that empowers users to easily create chipless smart textile interfaces. The toolkit features a web-based design environment for importing sewing patterns, placing sensing components, and defining the reading spot. It abstracts the complexity of textile interface and component design, enabling users to build a chipless interface without engaging in RF circuit theory or embroidery path planning. It also automatically generates fabrication files for embroidery and cutting machines to facilitate the making of circuits and sensors. In addition, our toolkit includes a reader and testing software to validate the functionality of prototypes. 

\subsection{Web-based Design Environment}
The web-based design environment serves as the front end of \projecttitle{} for creating chipless smart textile interfaces. It supports importing sewing patterns in standard vector or PDF formats and renders them as a 2D background, allowing textile sensors to be designed in direct relation to garment geometry and functional regions.
An interactive canvas is overlaid on the sewing pattern, enabling users to drag textile sensors and the reading spot from a sidebar library onto the layout, then adjust their position, rotation, and preferred connection direction through a configuration menu. The environment also supports defining keep-out zones on the pattern to prevent component placement in non-flat or restricted areas where curvature could affect electrical performance.
Once the layout is complete, users can trigger the “Generate Design” function, which runs the automated generation process and overlays the resulting interface design on the canvas for visual verification. After confirmation, the system exports the files needed for fabrication and testing: (1) an embroidery file for embroidered resonant circuits, (2) a cutting file for textile sensors, and (3) a configuration file for testing software.

\subsection{Smart Textile Interface Design Generation}

One core function of our toolkit is to generate component and circuit designs for chipless smart textile interfaces. This process involves three steps that transform the initial placement layout into a fabrication-ready design.

The first step is to assign operating frequencies to the reference circuit and each sensor circuit. Following prior work~\cite{WU_BIT}, we reserve 26–30 MHz for the reference circuit and start assigning sensor frequencies from 25 MHz downward. For each assignment, we begin with the sensor that has the smallest minimum capacitance or inductance, since it resonates at the highest frequency. We then calculate the frequency range needed to cover its sensing range, and reserve that range. \rev{For example, for a bending sensor whose capacitance ranges from 8 to 15~pF, we assign its initial resonance to 25~MHz. The system then calculates a textile inductor of approximately 5~$\mu$H, producing a resonance near 25~MHz at 8~pF and expecting it to shift to 18~MHz as the capacitance increases to 15~pF. We then repeat this process for the sensor with the next-smallest minimum component value in descending frequency order while maintaining a 1~MHz gap between adjacent resonance ranges.}

Once the frequency ranges are assigned, the second step is to generate circuit components and place them within the garment layout. For each sensor circuit and the reference circuit, the toolkit first computes the target inductance and capacitance values. Given these targets, it then generates the designs of textile inductors and capacitors by varying their geometric parameters according to the empirical models derived in Section 4. For component placement, the toolkit begins with user-specified locations of textile sensors and reading spots, after which accompanying components are positioned nearby relative to their associated sensors. A greedy optimization strategy iteratively refines these placements to maximize remaining available space while avoiding overlaps, maintaining required clearances (e.g., coils separated by at least their diameter \cite{MutualCoupling}, 1mm away from keep-out zones and sewing pattern boundaries), and minimizing routing length. The details of the optimization algorithm is described in Appendix~\ref{Component Optimization}.

\begin{figure*}[t]
    \centering
    \includegraphics[width=1\linewidth]{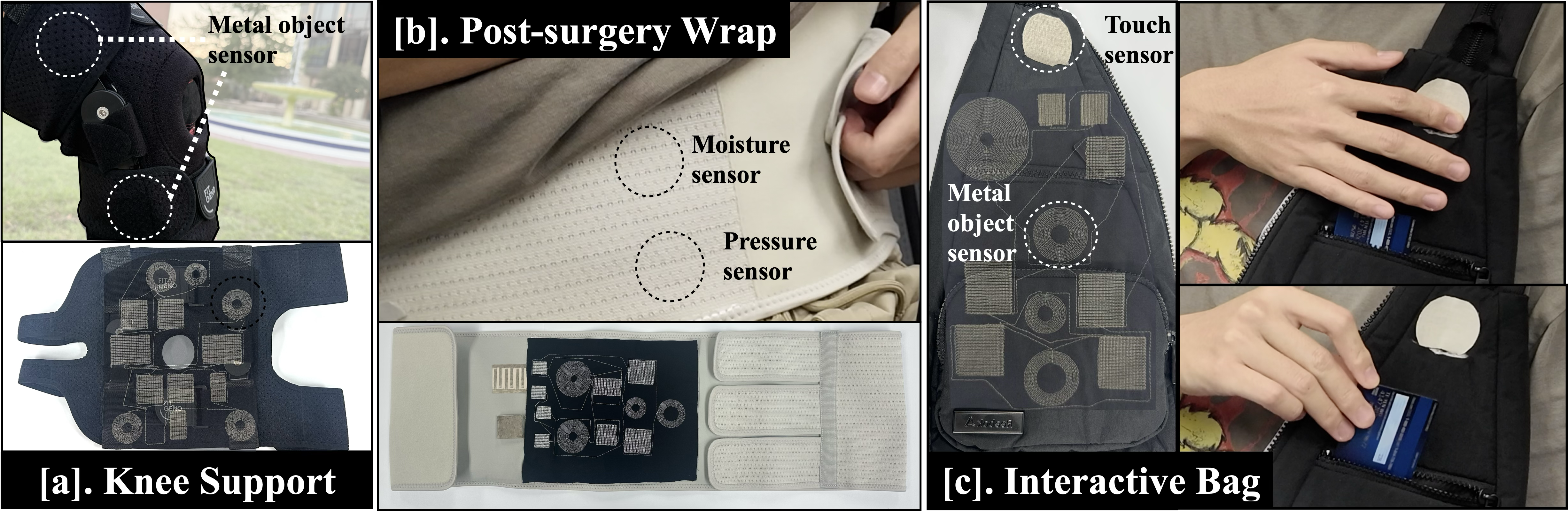}
    \caption{Three application examples created with \projecttitle{}: (a) a smart knee support that detects slipping of metal reinforcements; (b) a post-surgery wrap for monitoring wound status and wrap pressure; and (c) an interactive bag for detecting user interactions.}
    \label{fig:demo}
\end{figure*}

After component designs and positions are determined, the third step is to generate the complete circuit layout along with its transmission lines. This process uses a one-stroke embroidery, following the design rules in Section 5. This connects the receiver coil, reference circuit, and sensor circuits without requiring manual soldering. Transmission lines between sensor circuits are routed using an A* search algorithm, which identifies the shortest valid paths while respecting keep-out zones, maintaining minimum clearances, and staying within the garment’s usable textile regions. These steps finalize the generation of a chipless smart textile interface design.


\subsection{Fabrication and Configuration File}
Once the design is generated, our toolkit can prepare the files for fabrication and testing. Circuit layouts are exported as embroidery files (e.g., .PES) compatible with consumer embroidery machines, while cutting patterns are exported as SVG files for fabricating textile sensors and electrodes from conductive fabric. In our implementation, these files can be directly used with a Brother embroidery machine (Brother SE600 or NQ1700E) and a Cricut desktop cutter (Cricut Maker 3) to carry out circuit embroidery and sensor fabrication. Beyond fabrication, the toolkit also produces a system configuration file that specifies the operating frequencies, reference circuit parameters, and component details for each sensor circuit. This file can be loaded into the reader and testing software to facilitate prototype testing.

\subsection{Reader Hardware and Testing Software }
Our reader module serves as an external device responsible for wirelessly interrogating a chipless smart textile interface. It consists of a NanoVNA and a transmitter coil (outer diameter: 20 mm, inner diameter: 2 mm, turn gap: 1 mm, number of turns: 8, inductance: 0.472 µH) for transmitting excitation signals and receiving the impedance spectrum from the textile’s receiver coil. During testing, the reader connects to a computer via USB, where our custom software provides real-time data visualization. Before operation, the software performs calibration to fine-tune the capacitance of the SMA connector and the inductance of the transmitter coil. Users can then load a configuration file generated by our design tool, which automatically sets the operating frequencies and reference parameters for each sensor. During operation, the interface displays live sensor readings and continuously updates plots to reflect changes in capacitance or inductance in real time.

\section{APPLICATIONS}
To demonstrate the versatility of \projecttitle{}, we present three additional prototypes, beyond the smart shoulder brace described in Section 4, to showcase its capacity to support diverse chipless smart textile interfaces. These examples span different garment types and sensing needs, highlighting the toolkit’s potential for real-world applications. Each prototype only requires less than 50 cents in extra conductive wire and fabric.


\textbf{\textit{Knee Support:} }
For joint stabilization, knee supports often rely on rigid metal reinforcements. But sometimes it can shift out of place during movement, causing discomfort or even injury. To address this, we used \projecttitle{} to design a textile underlayer that is worn beneath the knee support. This layer integrates three object-detection sensors positioned to align with the two embedded metal reinforcements of the brace (Figure \ref{fig:demo}a). These sensors monitor whether the metal support moves significantly, offering insights on stability. Such functionality could provide wearers with alerts when adjustment is needed, improving both comfort and safety.

\textbf{\textit{Post-surgery Wrap:}}  
We used \projecttitle{} to design a smart post-surgery wrap for healthcare applications (Figure \ref{fig:demo}b). The wrap integrates a moisture sensor to detect wound exudate, providing a non-invasive means of assessing wound status. It also includes a pressure sensor to monitor whether the wrap is applied too loosely, which could slow recovery. When used on the chest or abdomen, the same pressure sensor can also capture respiratory activity by detecting subtle expansion and contraction of the body. These sensing capabilities enable continuous monitoring of patient conditions through fabric alone, supporting recovery in home settings.

\textbf{\textit{Interactive Bag:}  }
Finally, we created an interactive bag prototype with \projecttitle{} to demonstrate its potential for supporting subtle, everyday interactions (Figure \ref{fig:demo}c). A touch-sensitive sensor integrated into the top of the bag allows users to silence a smartphone’s ringtone with a simple tap, eliminating the need to locate and remove the phone. In addition, an object-detection sensor embedded in the front pocket monitors the presence of important belongings, such as credit cards or keys, providing reassurance before they leave a location.

\section{USER STUDY}  
The goal of our user study was twofold: (1) to evaluate the usability of the toolkit across different levels of user expertise, and (2) to collect qualitative feedback that could inform future improvements. 

\subsection{Participants}
We recruited 15 participants (N=15) with varying levels of expertise to evaluate \projecttitle{}. The cohort consisted of four professional designers\del{ (two from apparel design and two from art and design departments)}, three experienced makers with over one year of e-textile prototyping experience, and eight novices \del{from computer science and electrical engineering backgrounds}. This range allowed us to capture perspectives from experts, intermediate users, and beginners.

\subsection{Procedure}


During the study, we begin with a short introduction and demo of the toolkit, highlighting its core features such as the design canvas, component library, and overall fabrication process. Participants are then asked to use the toolkit to create smart textile prototypes. They are free to pursue their own ideas. If they do not have a specific project in mind, we suggest a smart shoulder brace prototype as a reference. \rev{Each participant completes at least one end-to-end design--fabrication--testing cycle.} During this process, participants explore the interface, place and configure sensing elements, and experiment with different layouts. After completing their design, they fabricate the prototype using the provided materials and test it with our software to verify functionality. Finally, participants fill out the System Usability Scale (SUS) questionnaire \cite{SUS} and take part in a semi-structured interview, where they reflect on the workflow, discuss challenges, and suggest features they would like to see in future versions of the toolkit.


\subsection{Results}

Overall, all 15 participants successfully completed the tasks\rev{ and made functional prototypes whose sensor readings responded to user interactions.} In the study, 5 participants followed the example to create a smart shoulder brace prototype, while the other 10 developed their own designs, including a smart diaper, a soft back sensor, a compression bra for mastectomy care, a prosthetic brace,a belt breathe band, a smart chair padding and a smart yoga sock. Images and detailed information of selected prototypes are shown in Appendix~\ref{Participant Design Prototypes}. The tool received a mean SUS score of 73.5 (SD = 6.3) across 15 participant, indicating acceptable usability \cite{SUS}, though there remains room for improvement. \rev{The common usability problems included insufficient guidance for selecting valid connection points, inadequate feedback when sewing-pattern regions were too constrained for circuit generation, limited explanation of keep-out zones, and ambiguity between the design schematic and the mirrored embroidered circuit, which could lead to sensors being attached in the wrong orientation.} In addition to them, we summarize three major findings from the qualitative feedback.



\textbf{\textit{Lowering Prototyping Barriers and Facilitating Workflow.}} Participants across different backgrounds reported Textro as reducing barriers in smart textile prototyping, though the specific reasons varied by experience. The four novice participants (P1, P3, P8, P12) emphasized the toolkit’s low cost and accessibility, explaining that they had previously perceived smart textile interfaces as inaccessible due to the need for electronics expertise. Participants with prior experience building wearable devices (P2, P4), by contrast, highlighted the difficulty in existing workflows when connecting conductive textile elements to sensing hardware , and felt that Textro made this process much easier. The fashion designer (P6, P7) offered a different perspective, explaining that current smart textile prototyping often depends on technicians, which slows iteration and limits opportunities for independent experimentation.


\begin{quote}
`` I usually need to rely on a technician to make the garment prototype functional, which slows things down and makes it harder to adjust the prototype on my own, even though it is a small change.'' - P6
\end{quote}

She noted that Textro could help streamline this process. These accounts suggest that Textro lowers not only technical barriers for novices, but also workflow barriers for more experienced designers.

\textbf{\textit{Encouraging Exploration, but Requiring More Guidance.} }
While participants were able to create prototypes with Textro, we observed that they often struggled to decide which sensors should be included in a garment and where they should be placed. For example, two novice participants were unsure which sensors would be appropriate for applications such as a smart shoulder brace for detecting arm rotation and elevation. One participant noted:

\begin{quote}
“I wasn’t fully sure whether I placed the sensor in the right location... I feel like I’d probably need a few more rounds of testing and adjustment once I see how it works in practice.” —P13
\end{quote}

This led users to begin with trial-and-error sensor placement and learn through repeated prototyping and testing. This suggests that, while Textro supports exploration, users still need to iteratively discover what works through experimentation. To better support users earlier in the design process, the system could provide more guidance \del{for both sensor selection and placement}. For instance, one participant mentioned that adding 3D visualization could help them map 2D sewing patterns to body locations more easily (P10). \rev{More advanced forms of design assistance, such as real-time previews of circuit layouts and routing, garment-aware keep-out-zone recommendations, and suggestions for suitable sensors and placements based on the intended application, could further reduce the cost of early experimentation and help users identify feasible designs before fabrication.}



\textbf{\textit{Testing Software Helps, but Sensor Interpretation and Application Remain Difficult.}}
While the testing software helped participants verify whether sensors were responsive and functioning, interpreting the sensor signals and relating them to specific body movements or interactions remained challenging. Participants could tell that a sensor changed, but they were often unsure how to understand that change, whether the response was meaningful, or how it could support their application. 
One participant noted:
\begin{quote}
``I could see the signal changing, but I wasn't really sure whether it actually corresponds to interaction I wanted to detect.'' - P9

\end{quote}
This suggests that, beyond basic testing, users would benefit from additional support for interpreting sensor behavior and connecting test results to application goals.
\section{DISCUSSION}

\textbf{\textit{Workflow Time and Iteration Cost.}}
\rev{Based on our user study, a typical \projecttitle{} workflow requires approximately 15~min for design, 30~min for embroidery and cutting, 10~min for assembly, up to 4~h for fabric-glue drying, and 10~min for reader alignment and signal readout. After producing an initial readable prototype, users typically complete two to five design--fabrication--testing cycles to refine the intended interaction or body placement. Each cycle consumes less than US\$1 of conductive thread and fabric, keeping the material cost relatively low. Although the cumulative duration can be substantial, much of it consists of passive drying time rather than active labor. Nevertheless, this delay interrupts the iteration cycle and postpones feedback on whether a design will perform as intended. Providing earlier assistance with sensor placement, circuit routing, and fabrication constraints could therefore reduce unsuccessful iterations, shorten the overall workflow, and lessen user frustration.}

\textbf{\textit{Interface-Level Sensing Performance.}}
\rev{
To preliminarily assess the sensing performance of a complete interface fabricated through our work, we conducted a lightweight technical validation using the smart shoulder-brace prototype. For the pressure sensor, we applied weights of 20, 50, 100, 150, and 200~g, with each condition repeated 5 times. For the bending sensors, we tested angles of 30$^\circ$, 60$^\circ$, and 90$^\circ$, with each condition likewise repeated 5 times. Under each condition, we compared the capacitance estimated through our system with the measured ground-truth capacitance. Across all sensing conditions, the prototype achieved an average estimation error of 6.6\% (SD = 2.6\%), suggesting that the fabricated chipless interface can achieve sensing performance comparable to that reported in prior work\cite{WU_BIT}. However, a larger controlled study is still needed to systematically quantify the interface-level performance.
}

\section{LIMITATIONS AND FUTURE WORK}
In this section, we discuss the limitations of Textro, along with potential directions for future work.



\textbf{\textit{Toward Scalable Fabrication.}}
Textro is currently designed for home-grade embroidery and cutting machines, enabling low-barrier rapid prototyping. However, this limits both precision and fabrication size. Small items such as gloves are difficult to make because densely packed sensors cannot be stitched too small without losing consistency, while larger interfaces may exceed the embroidery area and require manual repositioning, increasing labor and misalignment risk. To scale beyond prototyping, future work should explore fabrication methods with higher precision, larger working areas, and better consistency, such as industrial embroidery or roll-to-roll production. A key step will be developing a translation program that adapts Textro-generated designs from prototype-oriented fabrication workflows to manufacturing-oriented processes.

\textbf{\textit{Expanding Sensor Library.}}
One of our future work will expand Textro’s sensor library to support a wider range of sensing capabilities and applications. Beyond the current set of textile sensors, this will include modalities such as stretch~\cite{Vogl2017} or temperature sensing~\cite{10.1145/3613905.3648646}, enabling more diverse forms of on-body interaction and body-state sensing. In addition, resonant sensors will also be incorporated \cite{10.1145/3746059.3747733}. Unlike designs that rely on separate inductive and capacitive components, resonant sensors integrate both capacitance and inductance into a unified sensor structure. This can significantly reduce the space required for sensor implementation, which is especially valuable for compact garments or designs with many sensing points. Incorporating such sensors could help Textro support denser and more space-efficient smart textile interfaces.


\textbf{\textit{Material and E-Waste Considerations.}} Our current system relies on enameled copper wire, nickel/copper plated fabric, and cotton substrates for prototyping because they are accessible, easy to fabricate with, and stable in use. While our approach already avoids major sources of e waste in conventional smart textiles, such as batteries, chips, and rigid connectors, conductive wires and metal coated fabrics may still create waste at scale. In future work, we plan to explore more biodegradable or recyclable alternatives, such as bio based conductive threads \cite{EcoThreads} and recyclable conductive fabrics \cite{IntelliTex,Upcycling}. Because changing materials can also affect electrical behavior, including component values and impedance spectra, future work should systematically evaluate different material combinations to expand the design space and support diverse applications.

\section{CONCLUSION}
We present \projecttitle{}, a novel toolkit that lowers the barriers to designing, fabricating and testing chipless smart textile interfaces. By optimizing textile capacitors and inductors, developing a solderless fabrication pipeline, and creating a web-based design environment, the toolkit makes it possible to build passive, chipless smart textile prototypes with readily accessible materials and tools such as embroidery and cutting machines. Through four application examples, we demonstrate how \projecttitle{} supports the creation of functional wearable and interactive artifacts in different use contexts. By conducting a user study with 15 participants of varying expertise, we gather insights into its usability and potential for smart textile prototyping. Looking ahead, future work will focus on scaling toward manufacturing, expanding sensor library, especially with resonant sensors, and making it compatible with eco-friendly materials to increase the sustainability and reduce e-waste. 
\bibliographystyle{ACM-Reference-Format}
\bibliography{reference}










\appendix

\section{Textile Capacitor Experiment}\label{textile capacitor}

\subsection{Fabrication Parameters}\label{Capacitor Fabrication Parameters}

We fabricated nine types of textile capacitors by combining three designs with three fabrication methods, making five replicates of each at the same size (Figure \ref{fig:capacitor_design}).  Note that we particularly selected embroidery with bare wire (BEL 1985-ND, Digikey)  as one fabrication method because it potentially allows solderless connections by intersecting two wires.

\begin{figure}[h]
    \centering
    \includegraphics[width=1\linewidth]{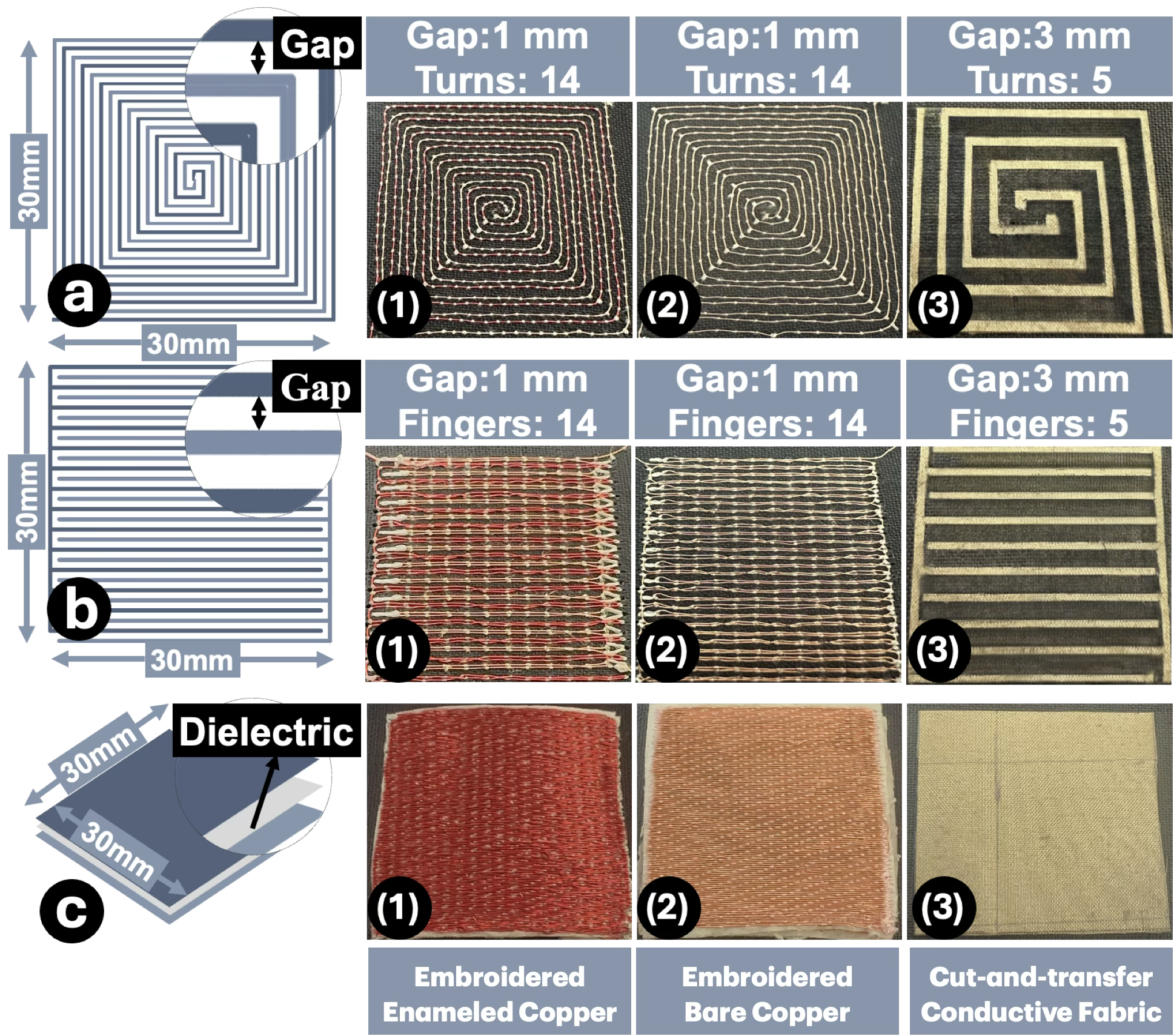}
    \caption{(Left) Capacitor designs evaluated in our experiments: (a) meander square, (b) interdigital, and (c) parallel-plate. Each design was fabricated using three methods: (1) embroidery with AWG 34 enameled copper wire, (2) embroidery with AWG 34 bare copper wire, and (3) cut-and-transfer with silver–nickel conductive fabric. Parameters and dimensions are shown in the figure. The substrate and dielectric layer of the parallel-plate conductive fabric capacitor are made of muslin cotton fabric. To ensure structural stability, all embroidered samples incorporate an embroidery stabilizer (Pellon 806 Stitch-N-Tear). The embroidery machine is Brother NQ1700E and the cutting machine is Cricut Maker 3.}
    
    \label{fig:capacitor_design}
\end{figure}

Fabrication parameters were optimized through a pilot study to establish a consistent baseline across different techniques. All components were designed with a standardized footprint of $30 \times 30$~mm. We implemented a 1~mm inter-electrode gap for embroidered samples and a 3~mm gap for cut-and-transfer samples. This configuration was strategically selected to balance structural stability, prevent electrical short circuits, and maximize component density.

For embroidered components, we utilized 34 AWG enameled copper thread (1175-1693-ND, Digikey) and 34AWG bare copper thread on a cotton fabric substrate (Springs Creative Black Solid Cotton Fabric), reinforced with an embroidery stabilizer (Pellon 806 Stitch-N-Tear) positioned beneath the fabric. This dual-layer stack preserved the integrity of the electrode patterns during high-speed stitching. To ensure high-fidelity reproduction, the thread tension of the Brother NQ1700e embroidery machine was calibrated to a value of 5. Precise tension control was paramount: excessive tension induced fabric puckering and distorted the component geometry, while insufficient tension resulted in loose stitch patterns and poor definition. These parameters were successfully validated on a secondary machine (Brother SE600), confirming the generalizability of our approach across consumer-grade hardware.

For the cut-and-transfer method, we employed a Cricut Maker 3 using mat mode cutting settings. We use Cricut StandardGrip Machine Mat as the rigid base for the machine to provide the stability. The conductive fabric (Adafruit Woven Conductive Fabric) was bonded to the cotton substrate using a thermo-adhesive layer (HeatnBond UltraHold Iron-on Clear Adhesive). TTo ensure high reproducibility and prevent electrical shorts, the inter-electrode gap for this method was set to 3 mm. This process ensured that the electrodes remained flush with the substrate, minimizing parasitic variations caused by air gaps or fabric shifting.

\subsection{Capacitor Evaluation Procedures}\label{Capacitor Evaluation Procedures}

\textbf{Measurement Setup} 
We employed a Vector Network Analyzer (NanoVNA) as the primary measurement instrument. Each capacitive component was interfaced via two 10~mm conductive leading lines soldered directly to the signal and ground pins of an SMA connector.

To isolate the intrinsic capacitance of the textile sensors, we implemented a calibration process using a parallel LC fitting model to compensate for parasitic effects introduced by the SMA connector and leading lines. Impedance spectra were collected across the Textro's operational frequency range of 5~MHz to 30~MHz. The parasitic values were computationally subtracted from the total measured impedance to yield the true component capacitance.


\textbf{Fabrication Error Evaluation.} The fabrication error was assessed using five replicates produced for each capacitor design. Following fabrication, the capacitance of each replicate was measured, and the Coefficient of Variation (CV) was calculated to quantify fabrication error.

\textbf{On-body Error Evaluation.} We evaluated the capacitor's on-body performance by attaching them to a volunteer's arm (Arm circumstance 28.2~cm, radius approximately 4.5~cm) using a velcro arm band, which ensures consistent contact pressure and placement across trials. The CV was calculated for on-body error to assess the stability of the textile-skin interface.

\textbf{Washability Evaluation.} To verify the structural and electrical durability of the sensors, we conducted a 10-cycle wash-and-dry stress test. The capacitors were enclosed in a mesh laundry bag (Brightroom Mesh Wash Bag) to simulate real-world garment care conditions. A Whirlpool Top Load Washer was used on a delicate mode (warm water, 15~ml detergent, 48~minute duration). Each wash cycle was immediately followed by a 90~minute drying cycle in a Whirlpool Vented Electric Dryer on a delicate mode, consistent with the standard care requirements for delicate clothing. After the 10th cycle, the capacitance was re-measured, and the CV was calculated to quantify the stability.

\subsection{Model Fitting for Interdigital Capacitor}\label{model fitting for Interdigital Capacitor}

\textbf{Measurement Setup.} The measurement setup procedures in the model fitting experiment were identical to those employed in the initial capacitor design evaluations.

\textbf{Empirical Fitting.} To predict the capacitance of an interdigital capacitor (IDC) with different numbers of fingers and varying lengths, our next step is to derive an empirical formula based on geometric parameters. To ground our model in established theory, we begin with the standard capacitance formula used for IDCs on printed circuit boards (PCBs). We then refine this formula using empirical measurements collected from our textile prototypes.

The standard formula for IDC design on PCBs is given by \cite{pozar_microwave_2012}:
\[
C_{\mathrm{pcb}} \;=\;
\frac{(\varepsilon_r + 1)}{W}\,
L \,\bigl[(N-3)\,a + b\bigr]
\quad\bigl[\text{pF}\bigr]
\]
where $L$ is the finger length, $W$ is the finger width, $\varepsilon_r$ is the relative permittivity, and $N$ is the number of fingers. The term fingers refers to the interleaving electrode tines. In this formulation, the finger gap is not included as a variable since it is fixed; in our design, the gap is 1~mm. The finger width is set to 0.32~mm, which corresponds to twice the diameter of the 34 AWG enameled wire due to the forward-and-back stitching path. The relative permittivity is approximated by the permittivity of the cotton substrate (1.6). To adapt this formula to our embroidered design, the parameters $a$ and $b$ must be derived through empirical fitting.

Since the model predicts that capacitance varies linearly with both the number of fingers and finger length (while all other factors are held constant), we focus our investigation on these two parameters. To isolate the effect of each factor, we adopt an orthogonal one-factor sweep design, systematically varying one variable while fixing the other. First, we fix the finger length at 40~mm and vary the number of fingers from 5 to 50 in increments of 5. Each design is fabricated with five replicas, and the resulting capacitance is measured using a NanoVNA. In the second sweep, we fix the number of fingers at 20 and vary the finger length from 5~mm to 50~mm in increments of 5~mm, again fabricating five replicas for each design and measuring their capacitances under the same setup. 

As a result, the baseline model parameters are estimated to be $a = 5.2$ and $b = 29.5$ ($R^2 = 0.99$). For on-body measurements, the parameters shift to $a = 5.8$ and $b = 43.6$ ($R^2=0.95$). 

\textbf{Substrate Material Verification.} To evaluate the generalizability of the proposed interdigital capacitor (IDC) formula, we conducted validation tests across a variety of fabric substrates, including silk (Fxomiev Pure Silk Fabric), polyester (Kitchen Ice Fabric Polyester), cotton (Springs Creative Black Solid Cotton Fabric), nylon (Rudong M Ripstop Nylon), linen (LOVOUS Nature Linen Needlework Fabric), and muslin (Akilens Wide Black Muslin). For each material, five replicates were fabricated using a consistent design ($L = 40$~mm, $n = 40$) to assess both prediction accuracy and fabrication consistency. The reliability of the sensor was quantified using the Coefficient of Variation ($CV = (\sigma/\mu) \times 100\%$), as summarized in Table~\ref{tab:substrate_test}. The results show that the formula maintains robust accuracy with a maximum relative error below $3\%$. Notably, Nylon exhibited the highest $CV$ ($2.78\%$), which can be attributed to its higher mechanical elasticity; varying material tension during the fabrication process likely introduced subtle geometric inconsistencies in the electrode spacing. However, its low average error ($2.00\%$) confirms the model's stability against such geometric variances. A key observation is that Cotton, despite being the fitting material, showed the highest error ($2.93\%$) while maintaining a low $CV$ ($1.39\%$). Given that the experiments were conducted under ambient room moisture conditions, this systematic shift is likely caused by the humidity-dependent permittivity. Unlike synthetic polyester (Error: $2.70\%$), cotton's effective permittivity is highly sensitive to micro-scale fluctuations in humidity between the fitting and testing phases. Conversely, the near-perfect fit on Muslin ($0.33\%$) and Silk ($1.07\%$) underscores the formula's superior accuracy on substrates with smoother surface morphologies and more stable dielectric properties. Overall, the consistently low error across both natural and synthetic textiles validates the model's generalizability for diverse wearable applications in real-world environments.

\begin{table*}[h]
\centering
\setlength{\tabcolsep}{5pt}
\caption{Measured Capacitance Replications vs. Predicted Values Across Different Fabric Substrates ($L=40$~mm, $n=40$)}
\label{tab:substrate_test}
\begin{tabular}{lcccccccc}
\hline
\textbf{Substrate} & \textbf{$R_1$ (pF)} & \textbf{$R_2$ (pF)} & \textbf{$R_3$ (pF)} & \textbf{$R_4$ (pF)} & \textbf{$R_5$ (pF)} & \textbf{Mean (pF)} & \textbf{CV (\%)} & \textbf{Error (\%)} \\ \hline
Silk               & 23.36 & 23.19 & 23.69 & 23.35 & 23.05 & 23.328 & 1.03\% & 1.07\% \\
Polyester          & 23.97 & 23.37 & 24.18 & 23.86 & 23.22 & 23.720 & 1.76\% & 2.70\% \\
Cotton             & 23.93 & 23.65 & 24.03 & 24.02 & 23.25 & 23.776 & 1.39\% & 2.93\% \\
Nylon              & 22.49 & 23.90 & 23.72 & 24.42 & 23.22 & 23.550 & 2.78\% & 2.00\% \\
Linen              & 23.34 & 22.77 & 24.03 & 23.25 & 23.65 & 23.408 & 2.01\% & 1.41\% \\
Muslin             & 23.39 & 23.03 & 22.97 & 22.34 & 23.28 & 23.002 & 1.78\% & 0.33\% \\ \hline
\end{tabular}
\end{table*}

\section{Textile Inductor Experiment}\label{textile inductor}

\subsection{Fabrication Parameters}\label{Inductor Fabrication Parameters}

The geometric layouts and dimensions for all designs evaluated in this study are illustrated in Figure~\ref{fig:inductor_design}. For the embroidered inductors, the gap between turns was fixed at 1 mm  for both enameled and bare copper thread, following the optimized parameters established in our pilot study. This consistency was maintained across all designs to ensure a stable baseline for comparison.

Both circular and square inductors were designed with a minimum inner diameter of 5 mm to accommodate the mechanical constraints of the embroidery machine. Our testing indicated that smaller diameters increase localized fabric tension, which can lead to significant structural deformation, specifically vertical and horizontal stretching, during the embroidery process. Maintaining this minimum diameter ensures the geometric fidelity of the conductive traces.

In the cut-and-transfer fabrication process, we maintained a 3 mm gap between the fingers of the meander inductor to ensure high reproducibility and prevent electrical shorts. To prevent material shifting or fraying during the cutting process, the conductive fabric was bonded to a cotton substrate using HeatnBond UltraHold Iron-on Clear Adhesive. Precision cutting was performed using a Cricut Maker 3 set to Mat Mode. A Cricut StandardGrip Machine Mat served as the rigid base to provide the necessary stability for intricate geometries.

All materials used for inductor fabrication, including substrates and conductive elements, remained consistent with those utilized in the capacitor experiments to ensure cross-component compatibility.

\begin{figure}[t]
    \centering
    \includegraphics[width=1\linewidth]{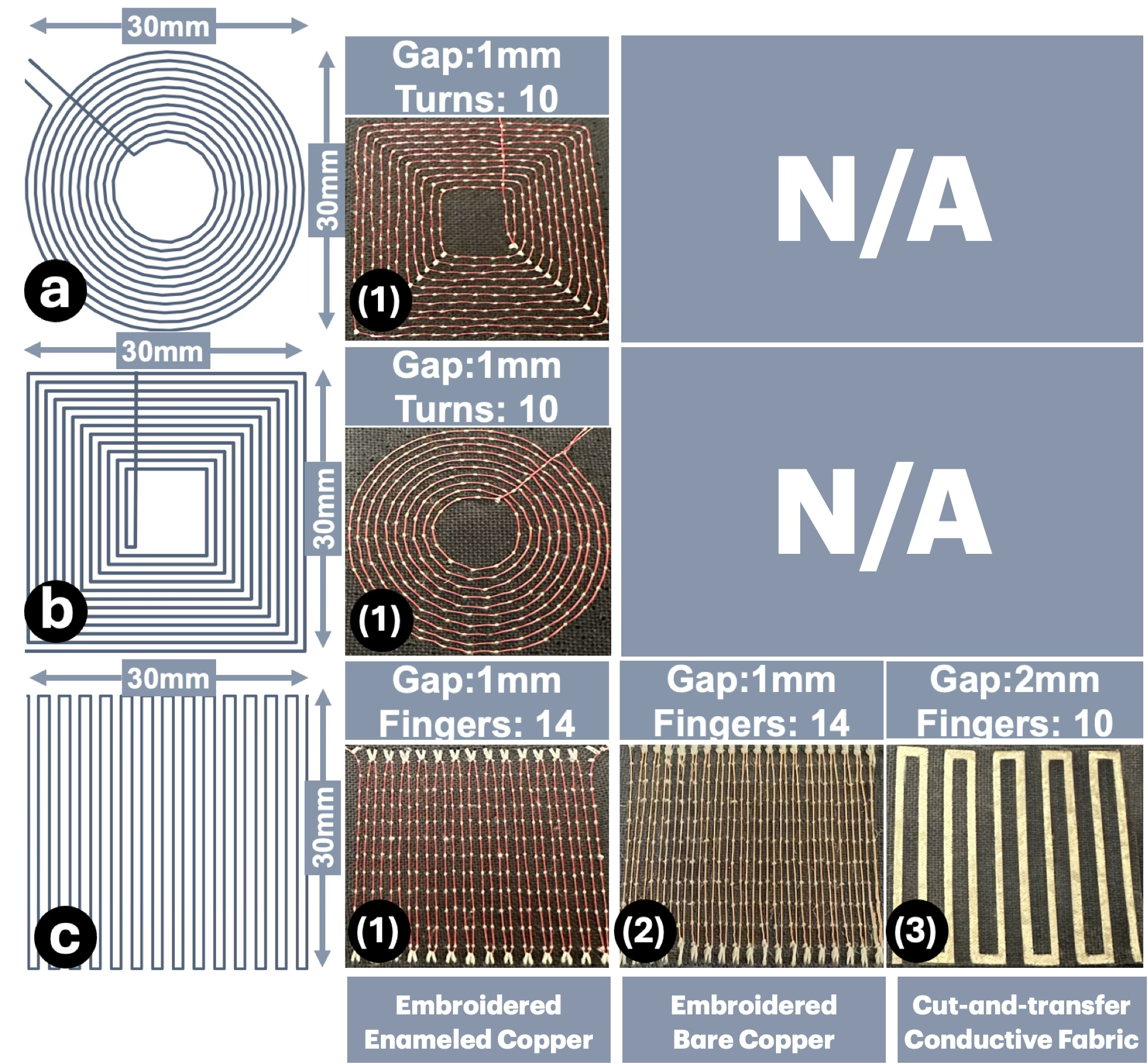}
    \caption{(Left) Inductor designs evaluated in our experiments: (a) square spiral, (b) circular spiral, and (c) meander line.
    We fabricated them using three methods: (1) embroidery with AWG 34 enameled copper wire, (2) embroidery with AWG 34 bare copper wire, and (3) cut-and-transfer with silver–nickel conductive fabric. Square and circular spirals were made only with enameled copper wire, as other methods caused short circuits. Fabrication parameters and dimensions are provided in the figure. The substrate is muslin cotton fabric.}
    \label{fig:inductor_design}
\end{figure}

\subsection{Model Fitting for Circular Inductor}\label{model fitting for circular inductor}
\textbf{Measurement Setup.} . The measurement setup used for model fitting is identical to that used in the initial capacitor evaluation. During calibration, the inductor is modeled as a series impedance together with the SMA connector and leading line.

\textbf{Empirical Fitting.} To predict the capacitance of an circular spiral inductor, we firstly introduced the Wheeler formula for inductance~\cite{Wheeler}:
\[
L_{\mathrm{pcb}} = \frac{0.001 \cdot N^2 \cdot D_{\text{avg}}^2}
         {a \cdot D_{\text{avg}} + b \cdot w},
\qquad
a \simeq 8,\; b \simeq 11
\]
To adapt this formula for textile fabrication, we also performed a one-factor parameter sweep to characterize the effect of geometry on inductance. Specifically, we vary the number of turns from 5 to 15 in steps of 2 while fixing the inner diameter at 12 mm. In addition, we vary the inner diameter from 5 mm to 25 mm in increments of 5 mm and fix the number of turns at 10. We calculate the average diameter (\(D\mathrm{avg}\)) by using the following formula:
\[
D_{\mathrm{out}} = D_{\mathrm{in}} + 2 N w, 
\qquad
D_{\mathrm{avg}} = \tfrac{1}{2}\,(D_{\mathrm{out}} + D_{\mathrm{in}})
\]

As a result, the textile inductance model parameters are estimated to be a=7.08, b=11 ($R^2=0.99$).

\section{Textile Sensor Design}\label{textile sensor design}

\subsection{Touch Sensor}

\textbf{Application.} Touch input is one of the most intuitive inputs for interacting with smart garments. The integrated touch sensor provides a flexible input interface, supporting both momentary and discrete button events. This enables seamless interaction with external systems, ranging from access control (e.g., automatic doors) to mobile and desktop computing interfaces.

\textbf{Fabrication Parameters.}
Our touch sensor comprises two adjacent, semicircular conductive fabric electrodes (Figure \ref{fig:sensor_design}a) \cite{10.1145/3654777.3676344, 10.1145/3379337.3415886}. The semicircular shape is chosen to highlight the interaction area while keeping the design compact. Conductive fabric is used because it provides a softer texture than embroidered electrodes. The conductive fabric is firstly cut into two semicircular shape in 2.54~cm radius, and then attached on the cotton fabric using iron-on adhesive and the spacing between two part is 2.54~mm. The sensor’s capacitance ranges from approximately 5 pF in the untouched state to 15 pF when touched with bare skin, offering a clear distinction between interaction states. In addition to binary touch input, the sensor supports proximity detection, capturing the user’s intent before physical contact occurs. For instance, the system can detect a hand approaching at a 1~cm range (marked by a shift from 5~pF to 9~pF), enabling pre-touch interactions such as triggering a visual preview, or waking up the device from a low-power state.

\subsection{Moisture Sensor}

\textbf{Application.} Moisture sensing is valuable in smart textile applications, especially in clinical contexts where maintaining a dry skin surface supports comfort and recovery~\cite{10.3389/fphy.2021.722173}. By tracking perspiration and localized humidity, these sensors provide real-time insights into skin conditions, facilitating the prevention of dermatological irritation and supporting the management of wound healing environments.

\textbf{Fabrication Parameters}
Our moisture sensor is implemented using an interdigitated capacitor design (Figure \ref{fig:sensor_design}b), where two sets of conductive fabric are patterned as comb-like thick fingers on a single fabric layer~\cite{6494581}. The sensor has an overall dimension of 2~cm by 3.81~cm and features three fingers per electrode. Each finger is designed with a width of 0.254~mm, separated by a 0.381~cm gap.

Then its bonded on the cotton fabric using iron on adhesive to leverage the hygroscopic properties of the cotton fabric. This structure increases the electrode surface area and enhances sensitivity to environmental changes. Built on a cotton substrate, the sensor shows a baseline capacitance of about 5 pF under dry, ambient conditions. As the fabric absorbs moisture, the dielectric constant increases, raising the capacitance to roughly 12 pF. 

\subsection{Pressure Sensor}

\textbf{Application.} Pressure sensing plays a vital role in smart textile systems,facilitating the measurement of biomechanical metrics such as posture, weight distribution, and joint articulation. These sensors enable continuous monitoring of muscle and joint activity during both dynamic exercise and sedentary behavior (e.g., sitting). Furthermore, this technology have the potential applications for clinical rehabilitation and assistive care, particularly in monitoring pressure points and mobility for patients with disabilities or limited motor function.

\textbf{Fabrication Parameters.}Our pressure sensor is built on a parallel-plate capacitor design (Figure \ref{fig:sensor_design}c), using two 3~cm by 3~cm conductive fabric electrodes separated by a 0.5~cm compressible sponge dielectric (i.e. polyurethane foam~\cite{9580528} and silicone elastomer~\cite{https://doi.org/10.1002/admt.201700237}). This arrangement allows the electrodes to move closer under applied force, directly translating pressure into measurable capacitance changes. At rest, the sensor maintains a baseline capacitance of about 6~pF. When pressed, the sponge compresses and the capacitance rises up to 18~pF, depending on the applied load. 

\subsection{Bending Sensor}
\textbf{Application.}Bending detection is widely used in smart textile interfaces, as it provides insight into body gestures and joint movements during daily activities and rehabilitation exercises. 

\textbf{Fabrication Parameters.} To address these needs, we develop a fabric-based bending sensor that measures directional deformation through a capacitive structure. Similar to our pressure sensor, it also adopts a parallel-plate design, but here the conductive fabric electrodes are arranged in a 1~cm by 4~cm rectangular form (Figure \ref{fig:sensor_design}d)~\cite{geissler2024embedding}. We use the 0.5~cm thick polyurethane foam as the dielectric layer. In its idle state, the sensor exhibits a baseline capacitance of approximately 8 pF. When the textile is flexed, the capacitance increases to as much as 15 pF, with the value proportional to the bending angle.

\subsection{Metal Object Detector}

\textbf{Application.}Object detection is a common sensing modality in HCI and healthcare applications, where the ability to track the presence and position of items is crucial. For example, in rehabilitation scenarios, clinicians need to monitor the placement of assistive accessories, such as braces, splints, or prosthetic attachments that are frequently made of metal. Proper alignment of these accessories is essential, as even misplacement can cause discomfort, reduce effectiveness, or impede recovery. 

\textbf{Fabrication Parameters.} To address this need, we develop coil-based inductive object sensors that can detect the presence and proximity of metal components. The sensor employs a wound spiral geometry (Figure \ref{fig:sensor_design}e) for direct embroidery. The sensor is approximately 40 mm × 40 mm, consisting of 13 turns and 25~mm inner diameter. In the idle state, the inductor exhibits an inductance of about 4.0 µH. When a metal object (e.g., a joint brace clip) approaches, the inductance decreases by roughly 1.1 µH, providing a clear signal for detecting presence and positional shifts.

\section{Requirement for Coupling Capacitance}\label{Requirement for coupling capacitance}

Let \(C\) denote the nominal capacitance used for the capacitive coupling connection and \(C_s\) the capacitance of the sensor.  
Given fabrication and environmental factors may vary the coupling capacitance; we model this as
\[
C' = C(1+\alpha),
\]
where \(\alpha\) is the relative variation.

In our coupling layout, two identical coupling capacitors are connected in series with the sensor (e.g., electrode--coupler, sensor, coupler--electrode).  
Thus, the effective capacitance seen by the reader is
\[
\frac{1}{C_{\mathrm{eq}}}
= \frac{1}{C_s} + \frac{1}{C(1+\alpha)} + \frac{1}{C(1+\alpha)}
= \frac{1}{C_s} + \frac{2}{C(1+\alpha)}.
\]

To quantify how coupling variation contaminates the sensor reading, we \emph{attribute} the effect of \(\alpha\) to an apparent sensor variation \(\delta\), while keeping the couplers at their nominal value \(C\):
\[
\frac{1}{C_{\mathrm{eq}}}
= \frac{1}{C_s(1+\delta)} + \frac{1}{C} + \frac{1}{C}
= \frac{1}{C_s(1+\delta)} + \frac{2}{C}.
\]
Equating the two expressions and letting \(k \triangleq C/C_s\),
\[
\frac{1}{C_s}\!\left(\frac{1}{1+\delta} + \frac{2}{k}\right)
= \frac{1}{C_s}\!\left(1 + \frac{2}{k(1+\alpha)}\right),
\]
which yields
\[
\frac{1}{1+\delta}
= 1 - \frac{2\alpha}{k(1+\alpha)}.
\]
Solving for \(\delta\) gives the exact attribution error as a function of \(k\) and \(\alpha\):
\begin{equation}
\boxed{\;
\delta
= \frac{2\alpha}{\,k(1+\alpha) - 2\alpha\,}
\;}.
\label{eq:delta_k_alpha}
\end{equation}

\paragraph{Design rule for a 10\% error budget.}
Suppose the acceptable sensor measurement error is \(\delta \le 10\%\) and the coupling capacitance may vary by \(\alpha = 20\%\). From \eqref{eq:delta_k_alpha},
\[
\delta = \frac{0.4}{\,1.2k - 0.4\,}
= \frac{1}{\,3k - 1\,}.
\]
Imposing \(\delta \le 0.10\) gives
\[
\frac{1}{3k-1} \le 0.10
\quad\Longrightarrow\quad
3k - 1 \ge 10
\quad\Longrightarrow\quad
\boxed{\;k \ge \tfrac{11}{3} \approx 3.67\;}.
\]
Thus, choosing the coupling capacitance at least \(4\times\) the sensor capacitance satisfies a 10\% error target under 20\% coupling variation. Using \(k=5\) (our standard choice) is a simple, robust rule-of-thumb:
\[
k=5 \;\Rightarrow\; \delta = \frac{1}{3\cdot 5 - 1} = \frac{1}{14} \approx 7.1\% \ (<10\%).
\]


\section{Component Placement Optimization}\label{Component Optimization}

We first determine the available placement regions based on the user-specified garment design and the keep-out zones. To support efficient placement, the algorithm decomposes the feasible space into a set of region trees, which represent candidate placement subregions after subtracting occupied areas and keep-out zones. This preprocessing step enables efficient recursive feasibility checks during iterative placement.

Patterns are separated into three groups: transmission lines, fixed-position patterns, and movable patterns. Transmission lines are excluded from this stage and are optimized later during routing. Fixed-position patterns, including textile sensors and the reader coil, remain at their predefined locations and are inserted directly into the occupancy map before movable components are placed.

The remaining movable patterns are grouped into four categories: coupling points, capacitors, and coils. These groups are processed in a predefined order, namely coupling points first, followed by capacitors and coils. This ordering prioritizes components that more strongly constrain subsequent connectivity and local layout. Within each category, patterns are ordered by their 2D footprint area in ascending order, such that larger components are placed before smaller ones.

Before recursive search, each component is assigned an initial anchor location based on its connection point and local branch structure. Specifically, the toolkit first derives an anchor from the component's geometric attributes (e.g., \texttt{top}, \texttt{left}, width, height, and connection point), and then instantiates initial component positions using fixed relative offsets from that anchor. For example, in the reference circuit, the left capacitor, reference inductor, and right capacitor are initialized at \((top-40, left-50)\), \((top-40, left)\), and \((top-40, left+50)\), respectively. Sensor branches follow the same principle, where the initial offsets are determined by the branch direction and gap parameters. These initialized positions provide the original placement used to guide subsequent local search.

For each movable pattern, the toolkit recursively traverses the region trees to identify feasible leaf regions that can accommodate the pattern footprint together with a safety margin. The search considers the component dimensions, margin requirement, the pattern's original location, and already placed patterns. In particular, existing patterns are included in feasibility checks to enforce spacing constraints, including those required for coils.

Each feasible leaf region yields a validated candidate \((x,y)\) placement for the component within that region, consistent with the component footprint, margin requirement, and spacing constraints. Thus, the search is carried out over discrete candidate placements induced by the decomposed free-space regions, rather than as a global continuous optimization. Candidate placements are then ranked using a lexicographic quality rule. When the pattern has an original placement, the algorithm prioritizes candidates with smaller displacement from the original position and uses region area as a secondary criterion. This favors placements that remain close to the initial design intent while also preferring larger feasible subregions. If no original placement is available, the ranking instead prioritizes larger feasible regions first and uses distance only as a secondary tie-breaker.

The highest-ranked feasible candidate is then selected greedily. The pattern is moved to the chosen location, and its dependent geometric information, such as connection points, is updated accordingly. The newly placed pattern is subsequently inserted into the occupancy representation by splitting all overlapping free-space regions across the region trees. This update ensures that later placement decisions only consider the remaining feasible space after earlier patterns have been committed.

If no feasible region is found for a movable pattern, the pattern is marked as unplaced. Overall, the component placement problem is solved using a sequential greedy heuristic rather than a global joint optimizer: the search space is defined by recursively traversed free-space region trees, candidate placements are ranked by displacement and region area, and the occupancy state is updated after each accepted placement so that subsequent decisions respect the remaining free space and spacing constraints.

\section{Participant Design Prototypes}\label{Participant Design Prototypes}

This section presents a diverse set of smart interface prototypes conceptualized and fabricated by the study participants. The participant cohort represents a multi-disciplinary group from various academic backgrounds and professional sectors, possessing a wide spectrum of technical expertise in wearable technology—ranging from novice makers to expert textile engineers. The resulting prototypes demonstrate the versatility of the wearable systems that could be made through \projecttitle{}. All designs were realized within a dedicated embroidery laboratory, where researchers provided the necessary conductive materials and fabric materials, allowing participants to translate their domain-specific insights into functional textile-based prototypes. 

\begin{figure*}[h]
    \centering
    \includegraphics[width=1\linewidth]{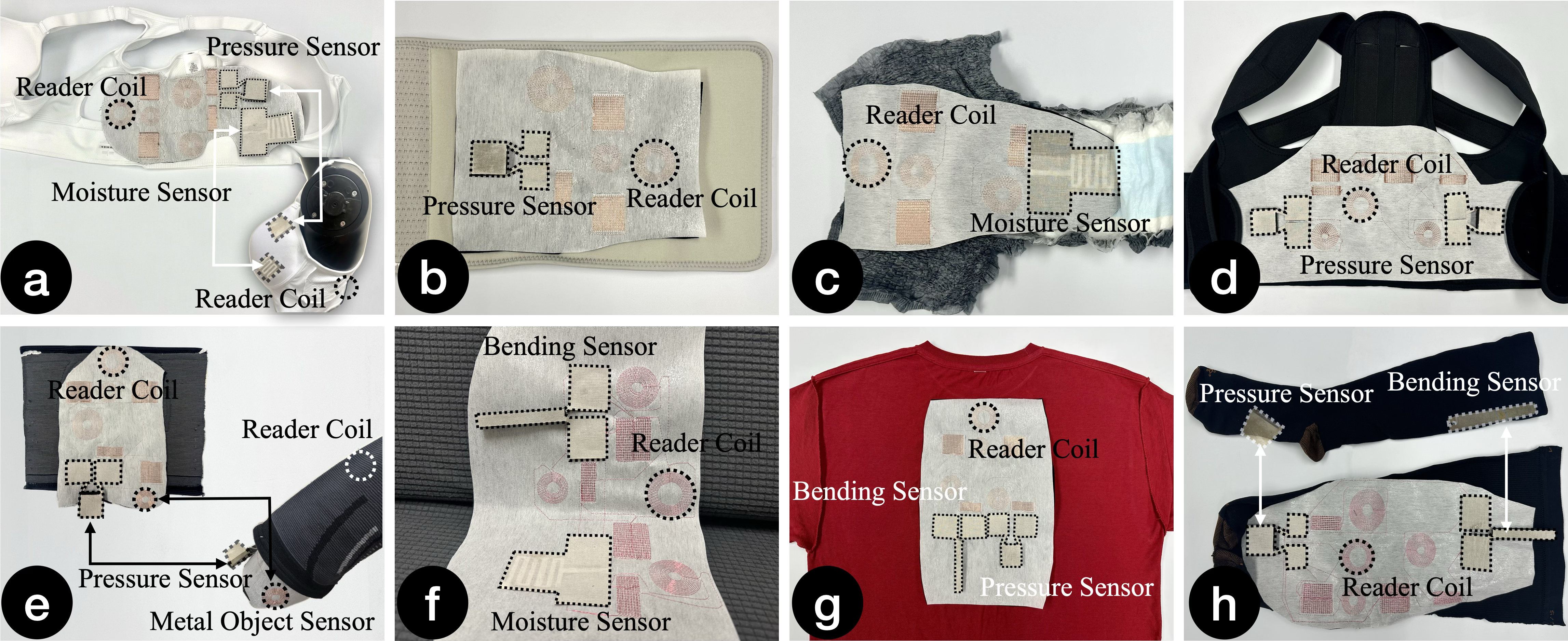}
    \caption{Selected smart textile interface prototypes designed and fabricated by participants: (a) A Sensing Bra for post-operative recovery monitoring; (b) A Smart Breathing Band tracking respiratory patterns via abdominal expansion; (c) A Smart Diaper for incontinence care and skin integrity; (d) A Back Strain Sensor on a brace for ergonomic posture assessment; (e) A Prosthetic Comfort Liner monitoring interface pressure and structural alignment; (f) Smart Chair Padding for monitoring sedentary duration and spinal alignment; (g) A Back Strain Sensing Shirt tracking lumbar muscle exertion; and (h) Smart Yoga Socks for postural correction and balance monitoring.}
    \label{fig:Participants_prototype_design}
\end{figure*}

\subsection{Sensing Bra}

The Sensing Bra was made by an apparel designer working at the intersection of textile engineering and women’s healthcare. This design specifically targets the management of post-operative wound and recovery following lumpectomy or mastectomy surgery. The garment integrates soft, fabric-based pressure and moisture sensors to monitor the surgical site. Continuous monitoring is critical because excessive pressure on the wound can impede localized blood circulation and lead to hematoma or lymphedema, while high moisture levels may cause skin maceration and significantly increase the risk of surgical site infections (SSIs), both of which can severely delay the healing process. The design demo is shown in Figure~\ref{fig:Participants_prototype_design}a.

\subsection{Smart Breathing Band}
The Smart Breathing Band was made by a novice designer exploring the integration of smart textiles into non-invasive physiological monitoring. This wearable device incorporates one pressure sensors to capture the biomechanical changes associated with respiration. When worn around the waist, the band detects the rhythmic abdominal expansion and contraction during the respiratory cycle. By correlating the localized pressure fluctuations with the curvature changes of the fabric, the smart band could track the user's respiratory rate and breathing patterns. The design demonstration of the Smart Breathing Band is illustrated in Figure~\ref{fig:Participants_prototype_design}a\b.

\subsection{Smart Diaper}
The Smart Diaper was made by a novice designer to address the critical biological challenges of long-term incontinence care. This design incorporates a fabric-integrated moisture sensor to continuously monitor the saturation level within the adult diaper, allowing caregivers to determine the optimal time for replacement. By maintaining the homeostasis of the skin’s microenvironment, the design aims to prevent maceration and Incontinence-Associated Dermatitis (IAD)—conditions caused by prolonged exposure to moisture levels. The design demo of the Smart Diaper, showcasing the seamless integration of the soft sensing textile to ensure wearer comfort and skin integrity, is illustrated in Figure~\ref{fig:Participants_prototype_design}c.

\subsection{Back Strain Sensor on Back Brace}
The Back Strain Sensor was made by a novice designer motivated by personal experiences with chronic back pain during prolonged sedentary work. This design integrates two soft fabric-based pressure sensors into a back support brace to facilitate real-time ergonomic assessment. The dual pressure sensors are positioned to monitor localized muscle exertion and load distribution across the lower back. By quantifying these biomechanical markers, they could detect poor sitting postures and provide feedback to mitigate the risk of long-term back strain. The demonstration of the integrated back brace is illustrated in Figure~\ref{fig:Participants_prototype_design}d.

\subsection{Prosthetic Comfort Liner}
The Prosthetic Comfort Liner was proposed by a novice designer . This design integrates one soft pressure sensor and one inductive (metal object) sensor to monitor the critical interface between the residual limb and the socket. The inductive sensor is utilized to ensure the structural alignment and correct positioning of the liner relative to the prosthetic’s internal metallic pylon or frame. Simultaneously, the pressure sensor monitors the mechanical load and interface pressure during ambulation. By combining these two sensing modalities, the liner could detect improper donning of the prosthesis and provide early warnings for excessive localized pressure, thereby preventing skin breakdown and improving overall user comfort. The demonstration of the smart liner interface is illustrated in Figure~\ref{fig:Participants_prototype_design}e.


\subsection{Smart Chair Padding}

The Smart Chair Padding was proposed by a maker with one-year smart textile prototyping experience. This design integrates a moisture sensor and a bending sensor directly into the contact surfaces of a standard office chair. The moisture sensor is embedded within the seat pan to monitor localized perspiration and heat accumulation, which serve as physiological indicators of prolonged occupancy. Meanwhile, the bending sensor is positioned along the backrest to capture real-time changes in the chair’s curvature, allowing the system to detect slouching or improper spinal alignment. By synthesizing these two data streams, the padding could provide feedback on both sitting posture and sedentary duration to promote healthier working habits. The demonstration of the instrumented chair padding is shown in Figure~\ref{fig:Participants_prototype_design}f.


\subsection{Back Strain Sensor on Shirt}
The Back Strain Sensor was made by an undergraduate maker with an interest in wearable technology and human-computer interaction. This design integrates soft fabric-based pressure and bending sensors directly into the posterior surface of a standard shirt. The sensors are parallel positioned at the lower lumbar regions of the back to monitor localized muscle exertion and spinal loading. By tracking the distribution of pressure across these key anatomical areas, the shirt could provide real-time feedback on muscle strain during various physical activities or prolonged sitting, helping to prevent musculoskeletal fatigue. The demonstration of the sensor-integrated shirt is illustrated in Figure~\ref{fig:Participants_prototype_design}g.

\subsection{Smart Yoga Sock}

The Smart Yoga Sock was made by a graduate student specializing in functional design. This prototype aims to facilitate postural correction and balance monitoring during yoga practice by integrating soft textile sensors into the sock's structure. A pressure sensor is embedded on the plantar surface of the foot to monitor weight distribution and center of pressure, while a bending sensor is positioned across the ankle to monitor the joint activities. By analyzing the synergy between foot pressure and ankle articulation, the smart sock could provide feedback on the stability and biomechanical alignment of complex yoga poses. The demonstration of the smart yoga sock is illustrated in Figure~\ref{fig:Participants_prototype_design}h.

\end{document}